      [1999/12/01 v1.4c Il Nuovo Cimento]
\documentclass{cimento}

\usepackage{graphicx}
\usepackage{amsmath}
\usepackage{bm}           
\usepackage{bbm}
\usepackage{color}
\usepackage{ulem}

\newcommand{\bra}[1]{\ensuremath{\langle#1|}}
\newcommand{\ket}[1]{\ensuremath{|#1\rangle}}
\newcommand{\Eins}{\ensuremath{\mathbbm 1}}
\newcommand{\mean}[1]{\ensuremath{\langle #1 \rangle}}
\newcommand{\ketbra}[1]{\ensuremath{| #1 \rangle \langle #1 |}}
\newcommand{\proj}[1]{\ketbra{#1}}
\newcommand{\tr}{{\rm Tr}}
\newcommand{\ud}{\mathrm{d}}
\newcommand{\be}{\begin{equation}}
\newcommand{\ee}{\end{equation}}
\newcommand{\beq}{\begin{eqnarray}}
\newcommand{\eeq}{\end{eqnarray}}
\newcommand{\POVM}{\hat{E}(\varepsilon)}
\newcommand{\POVMv}{\hat{E}(\vect{\varepsilon})}
\newcommand{\vect}[1]{\bm{#1}}

\title{Quantum theory of phase estimation}
\author{Luca Pezz\`e~\from{ins}  \atque
Augusto Smerzi~\from{ins}}
\instlist{\inst{ins} QSTAR, INO-CNR and LENS - Largo Enrico Fermi 6, Firenze, IT}
\PACSes{
\PACSit{42.50.St}       
{03.67.Bg}      
{03.75.Dg}      
{03.75.Gg}      
}

\begin{document}

\maketitle

\begin{abstract}
Advancements in physics are often motivated/accompanied by advancements
in our precision measurements abilities. The current generation of
atomic and optical interferometers is limited by shot noise, a fundamental limit
when estimating a phase shift with classical light or uncorrelated atoms. In the last
years, it has been clarified that the creation of special quantum correlations among
particles, which will be called here {\it useful entanglement}, can strongly enhance the
interferometric sensitivity. Pioneer experiments have already demonstrated the basic
principles. We are probably at the verge of a second quantum revolution where
quantum mechanics of many-body systems is exploited to overcome the limitations
of classical technologies. This review illustrates the deep connection between entanglement
and sub shot noise sensitivity.
\end{abstract}

\newpage

\tableofcontents


\section{Introduction}

\begin{figure}[b!]
\begin{center}
\includegraphics[clip,scale=4.5]{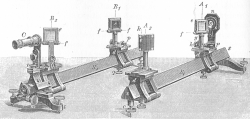}
\end{center}
\caption{{\bf Mach-Zehnder interferometer.} 
The Mach-Zehnder is the {\it drosophila} of two-mode interferometers. This is the picture taken from the original paper of Zehnder 
who was interested to measure the effect of pressure on the refractive index of water \cite{Zehnder}. The same apparatus was 
designed independently 
by Ludwig Mach to study nonstationary gas dynamics \cite{Mach}.} \label{Fig:MZ}
\end{figure}

Interferometry is the art of estimating phase shifts.
An interferometer is a physical apparatus that encodes the value of a parameter 
into a probe state.
In optical interferometers a phase shift is generally induced by a lapse in 
the relative time taken by the light to travel down two distinct paths (the interferometer arms).
This might probe the existence of eather, as in the first Michelson-Morley interferometer \cite{MM}, 
a supersonic airflow perturbing one optical path as in the first Mach-Zehnder \cite{Zehnder, Mach} (see Fig. \ref{Fig:MZ}),
or the occurrence of ripples in the curvature of spacetime, as in current $3$km long gravitational wave detectors 
(for a review see \cite{PitkinLRR2011}).
Differently from photons, atoms couple with inertial forces. 
This has prompted the development of atom interferometers (for a review see \cite{CroninRMP2009}) which
are very sensitive to accelerations and rotations \cite{InguscioBOOK}.
The current generation of atom interferometers nowadays reaches unprecedented precisions 
in the measurement of gravity \cite{PetersNATURE1999}, inertial forces
\cite{GustavsonPRL1997, GeigerNATCOMM2011}, atomic properties \cite{EkstromPRA1995} and fundamental constants \cite{FixlerSCIENCE2007,LamporesiPRL2008,BouchendiraPRL2011}. 
Moreover, if the two modes supporting the dynamics are two internal levels, 
the measurement of atomic transition frequencies with Ramsey interferometry \cite{RamseyBOOK} 
can be exploited for spectroscopic purposes in general and to determine a frequency standard for atomic clocks \cite{DiddamsSCIENCE2004}.

Phases can be estimated but cannot be measured. There is not, in quantum mechanics, such a thing like a Hermitian operator
corresponding to a quantum phase \cite{LynchPHYSREP1995}. 
In this sense, phases share the same fate of time, as both have to be considered as parameters in the quantum realm. 
The estimation of a phase shift is done by choosing an observable -- one having a corresponding Hermitian operator --
and decrypting the statistics of measurement results on the output state of the interferometer. 
The central goal of interferometry is to choose probe state, interferometric transformation, observable to be measured 
(typically, the number of particles measured at the output of the interferometer)
and, finally, estimator (a good one is the maximum likelihood), 
in order to infer the phase shift with the smallest possible error, given finite available resources.
What is this limit ? 
The phase $\theta$ estimation sensitivity of a two-modes interferometer probed by uncorrelated particles
is set by the shot noise limit $\Delta \theta_{\rm SN} = 1/\sqrt{Nm}$, where $N$ is the number of particles in the 
input state and $m$ is the number of independent measurements done with identical copies of the input.
This has been considered for a long time a fundamental limit
till Caves got the idea to squeeze the vacuum fluctuations entering in the unused port of the Michelson-Morley interferometer \cite{CavesPRD1981}. 
After this work, there have been several proposals \cite{BondurantPRD1984,Yurke_1986_b,Yurke_1986,YuenPRL1986,VourdasPRA1990,HollandPRL1993,
HilleryPRA1993,AgarwalPRA1994,SandersPRL1995,BrifPRA1996,JacobsonPRL1998,BollingerPRA1996,
DowlingPRA1998,BuzekPRL2000,DunninnghamPRL2002,CombesJOB2005,PezzePRA2006,UysPRA2007,
DurkinPRL2007,PezzePRL2008,CablePRL2010,JanPRA2010,JooPRL2011,PezzePRL2013}
on increasing the sensitivity above the shot noise, some of them explicitly
relating the sensitivity enhancement with the creation of some sort of quantum correlation
\cite{WinelandPRA1992, WinelandPRA1994, SorensenNATURE2001, DArianoPRL2001}. 
Only very recently it has been clearly shown that in order to overcome the shot noise it is necessary to generate 
entanglement among the particles of the input state \cite{GiovannettiPRL2006, PezzePRL2009,HyllusPRA2012,TothPRA2012}.
Entanglement -- the biggest mystery of quantum mechanics -- can increase the sensitivity of an interferometer beyond the shot noise up
to the Heisenberg limit $\Delta \theta_{\rm HL} = 1/N\sqrt{m}$ \cite{GiovannettiPRL2006,PezzePRL2009, HyllusPRA2012,TothPRA2012}. 
This theoretical prediction is the subject of an intense experimental activity with 
photons \cite{GrangierPRL1987, XiaoPRL1987, RarityPRL1990, KuzmichQSO1998, WaltherNATURE2004, MitchellNATURE2004, NagataSCIENCE2007, OkamotoNJP2008, SunEPL2008, AfekSCIENCE2010, XiangNATPHOT2010, KacprowiczNATPHOT2010, KrischekPRL2010}
-- with emphasis on the application to gravitational wave detectors 
\cite{McKanziePRL2002, VahlbruchPRL2005, GodaNATPHYS2008, VahlbruchPRL2008, SchnabelNATCOMM2010, LIGO2013} --
trapped ions \cite{MeyerPRL2001, LeibfriedSCIENCE2004, LeibfriedNATURE2005, RoosNATURE2006, GaoNATPHYS2010, MonzPRL2011}, 
cold \cite{HaldPRL1997, KuzmichPRL2000, AppelPNAS2009, Schleier-SmithPRL2010, LerouxPRL2010, WasilewskiPRL2010, Louchet-ChauvetNJP2010, ChenPRL2011, SewellPRL2012} and 
ultra-cold \cite{OrzelSCIENCE2001, EsteveNATURE2008, GrossNATURE2010, RiedelNATURE2010, LueckeSCIENCE2011, GrossNATURE2011, BuckerNATPHYS2011, 
ChapmanNATPHYS2012, BarradaNATCOMM2013, OckeloenPRL2013} atoms.
In particular, Bose-Einstein condensates, thanks to the large intrinsic nonlinearities due to particle-particle interaction, 
have attracted large interest \cite{SorensenNATURE2001,PezzePRL2009,
CiracPRA1998, GordonPRA1999, PuPRL2000, DuanPRL2000, HelmersonPRL2001, SorensenPRA2002, MicheliPRA2003, YouPRL2003, ZhangPRA2003, WeissPRL2008, LiuPRL2011} 
for the creation of entangled states of a large number of atoms.
However, not all entangled states can provide a sub shot noise phase sensitivity. Which entanglement
is really useful to overcome classical interferometry ? This was also discovered only a few years ago: 
the useful entanglement is the one, and only one, recognized by the Fisher information \cite{PezzePRL2009, HyllusPRA2012,TothPRA2012}. The recognition
and exploitation of entanglement useful for phase estimation sets the field of {\it quantum interferometry}. 

This review article is devoted on elaborating in some details the concept outlined above. But 
there is a last question which needs a quick answer: why bother about quantum interferometry~?  To one hand,
we of course know that leaps in physics are very often motivated/accompanied 
by leaps in our precision measurements abilities. 
Entanglement can boost phase sensitivities, providing the next generation of 
ultrasensitive device. To the other hand, quantum interferometers are 
fascinating toolboxes to learn about foundational questions of quantum mechanics. This review contribution is
eventually devoted to exploring what interferometry can tell us about it. 
There are in the literature several review papers covering different facets of quantum interferometry
\cite{Luis_2000, GiovannettiSCIENCE2003, DunninghamCONTPHYS2006, DowlingCONTPHYS2008, 
ParisIJQI2009, WisemanBOOK, GiovannettiNATPHOT2011, MaPHYSREP2012}. 
Our paper focuses on theoretical aspects of phase estimation, with special emphasis on the role played by entanglement.


\section{Phase Estimation}
\label{ClassEst}

How much precise can a statistical estimation be~?
Is there any fundamental limit~?
These are the central questions of the theory of statistical inference. 
The first answers came around 1940s with the works of 
Rao \cite{Rao1945}, Cram\'er \cite{CramerBOOK} and Fr\'echet \cite{Frechet1943} 
(extended to the multi parameter case by Darmois \cite{Darmois1945}), which independently 
found a lower bound to the variance of an arbitrary estimator.
This bound, generally indicated as the Cram\'er-Rao (lower) bound,
is intimately related to the Fisher information, introduced by Fisher in 1920s \cite{Fisher1922}~\footnote{
Besides the Cram\'er-Rao lower bound, different bounds of phase estimation have been 
introduced in the literature \cite{Bhattacharya, Barankin}. 
These are particularly relevant (stronger than the Cram\'er-Rao bound) in the non-asymptotic regime, 
i.e. for a small number of measurements.
However, they are rather difficult to calculate and will not be discussed in this review.
}. 
The Fisher information thus plays a central role in the theory of phase estimation. 
Its maximisation over all possible quantum measurements defines the so-called quantum Fisher information
\cite{BraunsteinPRL1994, BraunsteinANNPHYS1996} and provides a quantum lower limit to the 
Cram\'er-Rao bound \cite{HelstromPLA1967, HelstromBOOK, HolevoBOOK}.

In this section we introduce and demonstrate the Cram\'er-Rao lower bound and its
relation with different estimation protocols, like the maximum likelihood and the method of moments.
We also discuss some important properties of the Fisher information and the quantum Fisher information
(for recent reviews covering this topic see~\cite{ParisIJQI2009, WisemanBOOK}).

\begin{figure}[h!]
\begin{center} 
\includegraphics[clip,scale=0.75]{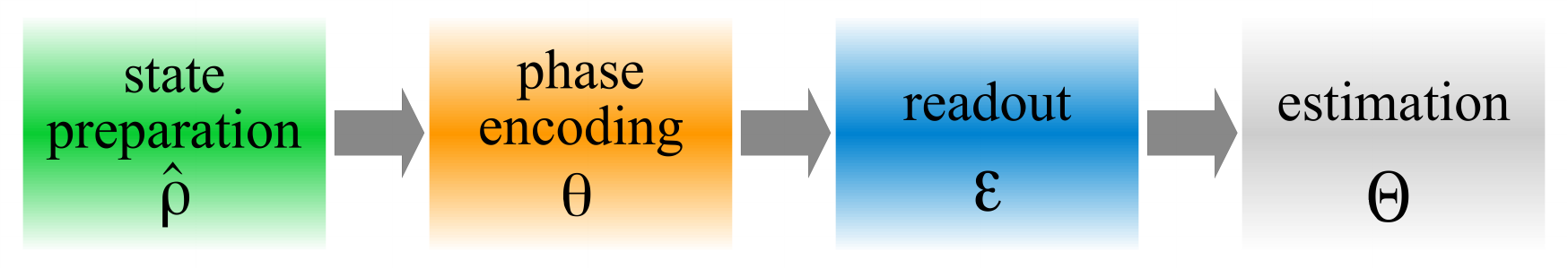}
\end{center}
\caption{Building blocks of phase estimation: $i$) the preparation of the probe state $\hat \rho$;
$ii$) the encoding of phase shift $\theta$, which transform the probe state to $\hat \rho(\theta)$;
$iii$) the readout measurement $\varepsilon$ and finally $iv$) the 
mapping from the measurement results to the phase provided by the estimator $\Theta(\varepsilon)$.
The phase sensitivity, i.e. the statistical variance of the estimator, depends crucially on all these operations.
} \label{Fig:scheme}
\end{figure}

\subsection{Basic concepts}
Here we fix the notation and introduce two basic concepts: {\it the likelihood function} and {\it the estimator}.\\
\\*
The building blocks of phase estimation are shown in Fig.~\ref{Fig:scheme}.
We consider an input (probe) state $\hat{\rho}$.
The interferometer is, in our language, {\it any} transformation of the probe which can be parametrized 
by a real (unknown) number $\theta$. The estimation of $\theta$ procees from 
the results of measurements performed on the output state $\hat{\rho}(\theta)$. 
These outcomes can be discrete as, for instance, the number of particles measured at the
output modes of a Mach-Zehnder interferometer (see Sec.~\ref{SU2}), or continuos as the spatial intensity of 
a double slit interference pattern~\cite{JanNJP2011,JanNJP2012}.
The most general formulation of a measurement in quantum theory is a positive-operator 
valued measure (POVM), i.e. a set of Hermitian operators
$\hat{E}(\varepsilon)$ which are non-negative (to guarantee non-negative probabilities) and
satisfy $\sum_\varepsilon \hat{E}(\varepsilon) = \Eins$ (to ensure normalization).
The standard projective (von-Neumann) measurement is a particular POVM where the operators 
$\hat{E}(\varepsilon)$ are orthogonal projectors, satisfying 
$\hat{E}(\varepsilon)\hat{E}(\varepsilon') = \hat{E}(\varepsilon) \delta(\varepsilon-\varepsilon')$.
The conditional probability to observe the result $\varepsilon$ for a given value of $\theta$ (also called ``likelihood" in the literature) is 
\be \label{probablity}
P(\varepsilon \vert \theta) = \tr\big[\hat{E}(\varepsilon) \hat{\rho}(\theta)\big].
\ee
The most general situation is to have correlated subsystems, described by $\hat \rho$, 
and perform $m$ correlated measurements, described by $\hat E(\vect{\varepsilon})$, where
 $\vect{\varepsilon} = \{\varepsilon_1, \varepsilon_2, ..., \varepsilon_m \}$.
 In this case, Eq.~(\ref{probablity}) extends to $P(\vect{\varepsilon} \vert \theta) = 
 \tr\big[\hat{E}(\vect{\varepsilon}) \hat{\rho}(\theta)\big]$.
If the probe state is 
made of $m$ (independent) {\it uncorrelated subsystems}, 
 \be \label{rhomprod}
\hat{\rho} = \hat{\rho}^{(1)} \otimes \hat{\rho}^{(2)} \otimes ... \otimes \hat{\rho}^{(m)},
\ee
and we perform {\it local operations} [such that 
$\hat{\rho}(\theta) = \hat{\rho}^{(1)}(\theta) \otimes \hat{\rho}^{(2)}(\theta) \otimes ... \otimes \hat{\rho}^{(m)}(\theta)$] and
{\it statistically independent} measurements,
\be \label{indepmeas}
\hat E(\vect{\varepsilon}) = \hat E^{(1)}(\varepsilon_1) \otimes \hat E^{(2)}(\varepsilon_2) \otimes ... \otimes \hat E^{(m)}(\varepsilon_m),
\ee 
then the likelihood function simply becomes the product of the single-measurement probabilities:
\be \label{likfuninde}
P(\vect{\varepsilon} \vert \theta) = \prod_{i=1}^{m} P_i(\varepsilon_i | \theta),
\ee
where $P_i(\varepsilon_i | \theta) = \tr\big[\hat{E}^{(i)}(\varepsilon_i) \hat{\rho}^{(i)}(\theta)\big]$.
In analytical manipulations is often convenient to work with the log-likelihood function
\be \label{llf}
L(\vect{\varepsilon} | \theta) \equiv \ln P(\vect{\varepsilon} | \theta)= \sum_{i=1}^{m} \ln P_i(\varepsilon_i | \theta),
\ee
where the right-side equality holds for independent measurements, i.e. $P(\vect{\varepsilon} \vert \theta)$ as in Eq.~(\ref{likfuninde}).
Given the set of outcomes $\vect{\varepsilon}$ of a random variable, 
the estimator $\Theta(\vect{\varepsilon})$ is any mapping from $\vect{\varepsilon}$ onto the parameter space.
In other words, the estimator is a generic function associating each set of measurement results with an estimation 
$\Theta$ of the phase. 
A relevant example of estimator is the maximum of the probability for a given set of results, 
i.e. the maximum likelihood (see Sec.~\ref{MaxLik}).
In practice, the estimator has to be carefully chosen so that $\Theta$ is as close as possible to the true, unknown, value 
$\theta$ of the phase. Since the estimator is a function of random outcomes, it is itself a random variable and can 
be characterized by its $\theta$-dependent mean value 
\be
\langle \Theta \rangle_\theta = \sum_{\vect{\varepsilon}} P(\vect{\varepsilon}\vert \theta)\,\Theta(\vect{\varepsilon})
\ee
(brackets $\langle ... \rangle$ indicate statistical averaging) and variance
\be
( \Delta \Theta)^2_\theta = \sum_{\vect{\varepsilon}} P(\vect{\varepsilon}\vert \theta)\,\big(\Theta(\vect{\varepsilon}) - \langle \Theta \rangle_\theta \big)^2.
\ee
It is clear that there are good choices and obvious bad choices of an estimator. 
The good choices are those which are unbiased (see below) and provide
the smallest uncertainty which, for almost all 
practical purposes, is quantified by the the square root of the variance.
More in details:

\begin{description}

\item{\it Unbiased estimators.} An estimator is said to be {\it unbiased} (for $m$ measurements) if its statistical average 
coincides with the true value of the parameter, 
\be
\langle \Theta(\vect{\varepsilon}) \rangle_\theta = \theta, \qquad \forall \, \theta, 
\ee
otherwise it is called biased.
In particular, for unbiased estimators we have 
$ \frac {\partial \langle \Theta \rangle_\theta}{ \partial \theta} = 1$.
An estimator which is unbiased only for certain values of the parameter is said to be 
{\it locally unbiased}~\footnote{More precisely, the estimator $\Theta$ is locally unbiased at the phase value 
$\theta_0$ if $\langle \Theta(\vect{\varepsilon}) \rangle_{\theta_0}=\theta_0$ and $ \frac {\partial \langle \Theta \rangle_\theta}{ \partial \theta} \big\vert_{\theta_0}= 1$.}.

\item{\it Consistent estimators.}
When performing a sequence of measurements $\vect{\varepsilon}=\{ \varepsilon_1,...,\varepsilon_m \}$, 
we can construct a sequence of estimates $\Theta(\varepsilon_1), \Theta(\varepsilon_1,\varepsilon_2), ..., \Theta(\varepsilon_1,\varepsilon_2,...,\varepsilon_m)$.
The estimator $\Theta(\vect{\varepsilon})$ is said to be {\it consistent} if such a 
sequence converges in probability to $\theta$. In other words, 
\be
\lim_{m \to \infty} {\rm Pr}\big( \vert \Theta(\vect{\varepsilon}) - \theta \vert > \delta \big) =0, \qquad \forall \, \theta, 
\ee
where $\delta$ is an arbitrary small number and $m$ is the sample size.
A consistent estimator is also 
{\it asymptotically unbiased}, 
\be
\lim_{m \to \infty} \langle \Theta(\vect{\varepsilon}) \rangle_\theta = \theta, \qquad \forall \, \theta.
\ee

\end{description}


\subsection{The Cram\'er-Rao lower bound and the Fisher information}
\label{crlb}

The Cram\'er-Rao is one of the most important theorems of phase estimation. It sets a lower bound to the variance 
of any arbitrary estimator: 
\begin{equation} \label{Eq:crlb}
(\Delta \Theta)^2_\theta \ge \big( \Delta \Theta_{\rm CR} \big)^2_\theta \equiv
{\big({{ {\partial \langle \Theta \rangle_\theta} \over { \partial \theta}}\big)^2 }
\over {I(\theta)}},
\end{equation}
where $ I(\theta)$ is the Fisher information (FI),
\be \label{MSFI}
I(\theta) \equiv \bigg\langle \bigg(\frac{\partial L(\vect{\varepsilon}\vert \theta) }{ \partial \theta}\bigg)^2 \bigg\rangle_\theta = 
\sum_{\vect{\varepsilon}} \frac{1}{P(\vect{\varepsilon} | \theta)} \bigg( \frac{\partial P(\vect{\varepsilon} | \theta)}{\partial \theta} \bigg)^2,
\ee
with the sum extending over all possible values of $\vect{\varepsilon}$~\footnote{
 In this Review we will always assume that the range of possible $\vect{\varepsilon}$ values does not depend on $\theta$. 
 As a consequence we have $\sum_{\vect{\varepsilon}} \frac{\partial P(\vect{\varepsilon} \vert \theta)}{\partial \theta} = \frac{\partial}{\partial \theta} \sum_{\vect{\varepsilon}} P(\vect{\varepsilon} \vert \theta) =0$,
 due to the normalisation condition $\sum_{\vect{\varepsilon}} P(\vect{\varepsilon} \vert \theta)=1$ which holds independently of $\theta$.
 In general, we can interchange the order of the differentiation over $\theta$ and the sum over $\vect{\varepsilon}$.
}. 
Equation~(\ref{Eq:crlb}) is the most general form of the Cram\'er Rao lower bound (CRLB). 
However, the bound is most useful when considered for unbiased estimators,
for which $ \frac {\partial \langle \Theta \rangle_\theta}{ \partial \theta} = 1$.
In this case the CRLB simply reduces to the inverse of the FI.
An estimator that saturates the CRLB is said to be {\it efficient}.
There is no guarantee that efficient estimators exist for an arbitrary number of measurements.
As shown below, the existence of efficient estimators depends on the properties of the probability distribution.
Nevertheless, in the limit of a large number of measurements, at least one efficient estimator exists:
the maximum likelihood estimator (see Sec.~\ref{MaxLik}).

The CRLB can be easily demonstrated. 
By combining 
\begin{eqnarray}
\label{crlb_2}
\frac{\partial \langle \Theta \rangle_\theta}{\partial \theta} = \frac{\partial }{\partial \theta} 
\sum_{\vect{\varepsilon}} \, P(\vect{\varepsilon }| \theta) \, \Theta(\vect{\varepsilon}) =
\sum_{\vect{\varepsilon}}
P(\vect{\varepsilon} | \theta) \, \Theta(\vect{\varepsilon}) \, \frac{\partial L(\vect{\varepsilon} | \theta)}{ \partial \theta} =
\bigg\langle \Theta \frac{\partial L }{ \partial \theta} \bigg\rangle_\theta
\end{eqnarray}
and
\begin{eqnarray}
\label{crlb_1}
\frac{\partial }{\partial \theta} \sum_{\vect{\varepsilon}} P(\vect{\varepsilon} | \theta) 
= \sum_{\vect{\varepsilon}} P(\vect{\varepsilon} | \theta) \, \frac{\partial L(\vect{\varepsilon} | \theta)}{ \partial \theta} =
\Big\langle \frac{\partial L}{ \partial \theta} \Big\rangle_\theta =0,
\end{eqnarray}
we obtain
\be \label{crlb_1b}
\bigg( \frac{\partial \langle \Theta \rangle_\theta}{\partial \theta} \bigg)^2 = 
\Big\langle \big(\Theta - \langle \Theta \rangle_\theta \big) ~ \frac{\partial L}{ \partial \theta} \Big\rangle^2_\theta.
\ee 
The next step, which brings us directly to the CRLB, is to exploit the Cauchy-Schwartz inequality
$ \langle A^2 \rangle_\theta  \langle B^2 \rangle_\theta \geq \langle  A B \rangle^2_\theta$, where 
$A$ and $B$ are arbitrary real functions of $\vect{\varepsilon}$ and the equality is obtained
if and only if $B = \lambda A$, with $\lambda$ independent of $\vect{\varepsilon}$.
Taking $A=\Theta - \langle \Theta \rangle_\theta$ and $B=\partial L/\partial \theta$, we have
\begin{eqnarray}
\label{crlb_3}
\Big\langle \big(\Theta - \langle \Theta \rangle_\theta \big)^2 \Big\rangle_\theta \, 
\Big\langle \Big(\frac{\partial L}{ \partial \theta}\Big)^2 \Big\rangle_\theta \geq 
\bigg( \frac{\partial \langle \Theta \rangle_\theta}{\partial \theta} \bigg)^2, 
\end{eqnarray}
and thus recover Eq.~(\ref{Eq:crlb}) since $( \Delta \Theta )^2_\theta = \langle (\Theta - \langle \Theta \rangle_\theta)^2 \rangle_\theta$.
An estimator $\Theta(\vect{\varepsilon})$ saturates the CRLB at the phase value $\theta$ 
if and only if the Cauchy-Schwarz inequality (\ref{crlb_3}) is saturated,
i.e. if and only if  
\be \label{crlb_saturation}
\frac{\partial L( \vect{\varepsilon} \vert \theta)}{ \partial \theta} = 
\lambda_\theta \, \big(\Theta(\vect{\varepsilon}) - \langle \Theta \rangle_\theta \big), \qquad \forall \, \vect{\varepsilon},
\ee
where $\lambda_\theta = I(\theta) / \frac{\partial \langle \Theta \rangle_\theta}{\partial \theta}$.  
Unfortunately this condition is generally not quite helpful to construct an efficient estimator.


\subsubsection{Upper bound: the quantum Fisher information}
\label{QFI}

-- Here we derive an upper bound to the FI. 
This is obtained by maximizing the FI over all possible POVMs,
\begin{equation} \label{QFIdef}
I_Q\big[ \hat \rho (\theta) \big] \equiv  \max_{ \{ \hat{E}(\vect{\varepsilon}) \} } I\big[ \hat \rho (\theta), \{ \hat E(\vect{\varepsilon}) \} \big],
\ee
generally indicated as the quantum Fisher information (QFI)~\footnote{
In this Review we will use different notations for the FI and QFI, depending on the context. 
According to Eq.~(\ref{MSFI}), the FI is a function of $\theta$ and we indicate it as $I(\theta)$ (or $F(\theta)$ for the single measurement case).
However, the probability distribution depends on  the output state $\hat \rho(\theta)$ of the interferometer and the
POVM $\{ \hat E(\varepsilon) \}$ used. 
In some cases, to be more precise we use the notation
$I[\hat \rho(\theta), \{ \hat E(\vect{\varepsilon}) \}]$ (or $F[\hat \rho(\theta), \{ \hat E(\varepsilon) \}]$ for the single measurement case).
}.
In the following we show that \cite{BraunsteinPRL1994}
\be \label{QFIL}
I_Q\big[ \hat \rho (\theta) \big] = \tr\big[ \hat{\rho}(\theta) \hat L^2_{\theta} \big],
\ee
where the Hermitian operator $\hat L_{\theta}$, called
symmetric logarithmic derivative (SLD)  \cite{HelstromPLA1967}, is defined as the solution of the equation 
\begin{equation} \label{SLDdefinition}
\frac{ \partial \hat{\rho}(\theta)}{\partial \theta} = 
\frac{ \hat{\rho}(\theta) \hat L_{\theta} + \hat L_{\theta} \hat{\rho}(\theta)}{2}.
\end{equation}
From Eqs.~(\ref{Eq:crlb}) and~(\ref{QFIdef}) we thus have the chain of inequalities:
\be
\big( \Delta \Theta \big)^2_\theta \geq \big( \Delta \Theta_{\rm CR}  \big)^2_\theta \geq \big( \Delta \Theta_{\rm QCR}  \big)^2_\theta,
\ee  
where  
\be
(\Delta \Theta_{\rm QCR})^2_\theta \equiv \frac{ \big( \frac{\partial \langle \Theta \rangle}{ \partial \theta}\big)^2 }
{I_Q\big[ \hat \rho (\theta) \big]}
\ee
 is the quantum Cram\'er-Rao (or Helstrom) bound \cite{HelstromPLA1967}.

Let us now demonstrate Eq.~(\ref{QFIL}).
We start from the definition of FI, Eq.~(\ref{MSFI}), with 
$P(\vect{\varepsilon} | \theta) = \tr\big[\POVMv \hat{\rho}(\theta) \big]$
and its derivative
$\partial_\theta P(\vect{\varepsilon} | \theta) = \tr\big[ \POVMv \partial_\theta \hat{\rho}(\theta) \big]$:
\begin{eqnarray} \label{FisherInfo}
 I\big[ \hat \rho (\theta), \{ \hat E(\vect{\varepsilon}) \} \big]
&=& \sum_{\vect{\varepsilon}}  \, \frac{ \tr[\POVMv \partial_\theta \hat{\rho}(\theta) ]^2}
{ \tr\big[\POVMv \hat{\rho}(\theta) \big] }.
\end{eqnarray}
Using the definition of SLD, Eq.~(\ref{SLDdefinition}), we have
\begin{equation} \label{realP}
\tr\big[\POVMv \partial_\theta \hat{\rho}(\theta) \big] 
= \Re \big( \tr[ \hat{\rho}(\theta) \hat L_{\theta} \POVMv] \big),
\ee
where $\Re(x)$ and $\Im(x)$ are the real and imaginary part of the complex number $x$, respectively.
To derive Eq.~(\ref{realP}) we have used 
$ \tr[ \hat L_{\theta} \hat{\rho}(\theta) \POVMv] = \tr[ \hat{\rho}(\theta) \hat L_{\theta} \POVMv]^*$, 
which follows from the cyclic properties of the trace and the Hermiticity of the operators.
The bound to the FI is obtained by using the chain of inequalities
\beq \label{ineqQFI}
\Re \big( \tr[ \hat{\rho}(\theta) \hat L_{\theta} \POVMv] \big)^2 
\leq 
\big\vert \tr[ \hat{\rho}(\theta) \hat L_{\theta} \POVMv] \big\vert^2  
\leq 
\tr\big[\hat{\rho}(\theta) \POVMv \big] \tr\big[ \POVMv  \hat L_{\theta} \hat{\rho}(\theta) \hat L_{\theta} \big], \nonumber \\
\eeq
valid for all values of $\vect{\varepsilon}$.
The first inequality comes from $\Re(x)^2 = \vert x \vert^2 - \Im(x)^2 \leq \vert x \vert^2$,
with equality if and only if
\be \label{condition1}
\Im \big( \tr[ \hat{\rho}(\theta) \hat L_{\theta} \POVMv] \big) = 0, \qquad \, \forall \vect{\vect{\varepsilon}}.
\ee
The second inequality is due to Cauchy-Schwarz
\footnote{
We use the Cauchy-Schwarz inequality 
$\vert \tr[\hat{A}^{\dag} \hat{B}]\vert^2 \leq \tr[\hat{A}^{\dag} \hat{A}] \tr[\hat{B}^{\dag} \hat{B}]$,
with equality if and only if $\hat{A} = \lambda \hat{B}$, where $\lambda$ is a complex number.
We take $\hat{A} = \sqrt{\hat{\rho}(\theta)} \sqrt{\POVMv} $ and 
$\hat{B} =  \sqrt{\hat{\rho}(\theta)} \hat L_{\theta} \sqrt{\POVMv}$,
noticing that $\hat{\rho}(\theta)$ and $\POVMv$ are positive operators. 
The equality is thus obtained if and only if
$\sqrt{\hat{\rho}(\theta)} \sqrt{\POVMv} = \lambda_{\theta, \vect{\varepsilon}} \sqrt{\hat{\rho}(\theta)} \hat L_{\theta} \sqrt{\POVMv} $,
for all $\vect{\varepsilon}$ at fixed $\theta$.
} and is saturated if and only if 
\be \label{condition2}
\hat{\rho}(\theta) \big( \Eins - \lambda_{\theta, \vect{\varepsilon}} \hat L_{\theta} \big) \POVMv = 0, \qquad \forall \, \vect{\varepsilon},
\ee
where $\lambda_{\theta, \vect{\varepsilon}} = \tr[\hat{\rho}(\theta) \POVMv]/\tr[ \hat{\rho}(\theta) \hat L_{\theta} \POVMv]$.
Combining Eqs.~(\ref{realP}) and~(\ref{ineqQFI}) we obtain
\be 
\frac{ \big( \tr[\POVMv \partial_\theta \hat{\rho}(\theta) ]\big)^2}
{ \tr\big[\hat{\rho}(\theta) \POVMv \big] } \leq 
\tr\big[ \POVMv  \hat L_{\theta} \hat{\rho}(\theta) \hat L_{\theta} \big],  \qquad \, \forall \vect{\vect{\varepsilon}}.
\ee
Finally, summing over $\vect{\varepsilon}$ and using $\sum_{\vect{\varepsilon}} \POVMv = \Eins$ gives 
\beq \label{ineqQFI3} 
I\big[ \hat \rho (\theta), \{ \hat E(\vect{\varepsilon}) \} \big] \leq \sum_{\vect{\varepsilon}} 
\tr\big[ \POVMv  \hat L_{\theta} \hat{\rho}(\theta) \hat L_{\theta} \big] = \tr\big[ \hat{\rho}(\theta) \hat L^2_{\theta} \big].
\eeq
Interestingly, the right hand side of this equation does not depend on the POVM.
We can thus interpret this results as a maximisation of $I [ \hat \rho (\theta), \{ \hat E(\vect{\varepsilon}) \} ]$ over all possible POVMs only if 
there exists at least one POVM such that both Eqs.~(\ref{condition1}) and (\ref{condition2}) are fulfilled, namely, 
that the upper bound Eq.~(\ref{ineqQFI3}) can be saturated.
If $\hat \rho(\theta)$ is invertible, Eqs.~(\ref{condition1}) and (\ref{condition2}) are 
equivalent to $( \Eins - \lambda_{\theta, \vect{\varepsilon}} \hat L_{\theta}) \POVMv = 0$, $\forall\,\vect{\varepsilon}$, 
with $\lambda_{\theta, \vect{\varepsilon}}$ real~\footnote{
If $\hat{\rho}(\theta)$ is non-invertible, the equation $( \Eins - \lambda_{\theta, \vect{\varepsilon}} \hat L_{\theta} ) \POVMv = 0$ is
sufficient but not necessary to fulfill Eq.~(\ref{condition2}).
}.
Since $\hat L_{\theta}$ is a Hermitian operator, there is a complete set of states $\{ \ket{\varphi_l} \}$ such that 
$\hat L_{\theta} \ket{\varphi_l} = \gamma_l \ket{\varphi_l}$, with $\gamma_l$ real numbers.
Equation (\ref{condition2}) is fulfilled by choosing the POVM
made of projectors $\{ \hat E_l \} = \{ \ket{\varphi_l} \bra{\varphi_l}\}$ into the basis 
that diagonalises $\hat L_{\theta}$ and taking $\lambda_l = 1/\gamma_l$ \cite{BraunsteinPRL1994}.
With this choice, Eq.~(\ref{condition1}) is also satisfied since 
$\tr[ \hat{\rho}(\theta) \hat L_{\theta} \hat E_l] = \gamma_l \bra{\varphi_l} \hat{\rho}(\theta) \ket{\varphi_l}$ 
has an imaginary part equal to zero.
The proof is now complete: there exists at least one optimal measurement, which is the Hermitian observable
build with the orthonormal eigenstates of $\hat L_{\theta}$, such that
$I[ \hat \rho (\theta), \{ \hat E_l \} ] = I_Q [ \hat \rho (\theta) ]=\tr[ \hat{\rho}(\theta) \hat L^2_{\theta}]$.

It is worth to point out that the optimal POVM for which the FI is equal to the QFI
depends, in general, on $\theta$.
This poses the unpleasant issue that, in principle, we need to know $\theta$ in order to 
choose the optimal POVM.
An adaptive scheme to overcome this problem has been suggested in Ref.~\cite{Barndorff-NielsenJPA2000}
showing that the QFI can be attained asymptotically in the number of measurements, without any knowledge of the parameter. 

 
 \subsubsection{Convexity}
\label{convexityf}

-- Let us consider the state $\hat{\rho}(\theta) = \sum_k \gamma_k \hat{\rho}_k(\theta)$ with $\gamma_k>0$ and $\sum_k \gamma_k=1$.
We have $\tr[\hat{E}(\vect{\varepsilon}) \hat{\rho}(\theta)] = \sum_k \gamma_k \tr[\hat{E}(\vect{\varepsilon}) \hat{\rho}_k(\theta)]$ and,
equivalently,
\begin{eqnarray} \label{ConvProb}
P(\vect{\varepsilon}|\theta) = \sum_k \gamma_k P_k(\vect{\varepsilon}|\theta). 
\end{eqnarray}
It is possible to demonstrate~\footnote{
The convexity of the FI has been first proved in \cite{CohenIEEE1968}. 
Here we report a slightly simpler proof based on a Cauchy-Schwarz inequality. Using Eq.~(\ref{ConvProb}) we have
\beq
\bigg(\frac{\partial P(\vect{\varepsilon}|\theta)}{\partial \theta}\bigg)^2 =
\bigg( \sum_k \gamma_k \frac{\partial P_k(\vect{\varepsilon}|\theta) }{\partial \theta} \bigg)^2 
\leq 
\sum_k \gamma_k P_k(\vect{\varepsilon}|\theta) \, 
\sum_k \gamma_k \frac{1}{P_k(\vect{\varepsilon}|\theta)} \bigg( \frac{\partial P_k(\vect{\varepsilon}|\theta) }{\partial \theta} \bigg)^2. \nonumber
\eeq
Therefore, 
\beq
\frac{1}{P(\vect{\varepsilon}|\theta)} \bigg(\frac{\partial P(\vect{\varepsilon}|\theta)}{\partial \theta}\bigg)^2 
\leq \sum_k \gamma_k \frac{1}{P_k(\vect{\varepsilon} |\theta)} \bigg( \frac{\partial P_k(\vect{\varepsilon} |\theta) }{\partial \theta} \bigg)^2, \qquad \forall \vect{\varepsilon},\theta. \nonumber
\eeq
After summing both members of this inequality 
over $\vect{\varepsilon}$ we obtain Eq.~(\ref{crlb_5b}), which holds for all possible POVMs and values of $\theta$.
} that
\begin{eqnarray}
\label{crlb_5b}
I(\theta) \le \sum_k \gamma_k I_k(\theta),
\end{eqnarray}
where $I_k(\theta) \equiv \sum_{\vect{\varepsilon}} \frac{1}{P_k(\vect{\varepsilon}|\theta)}\big( \frac{d P_k(\vect{\varepsilon}|\theta)}{d \theta} \big)^2$.
Equation (\ref{crlb_5b}) is known as the convexity of the FI.

The QFI is also convex~\footnote{
The basic idea is to recognise that the QFI is a FI calculated for the POVM, 
$\{ \hat E_{\rm opt, \hat \rho(\theta)}(\varepsilon) \}$, which is optimal for the specific state $\hat{\rho}(\theta)$.
Taking $\hat{\rho}(\theta) = \sum_k \gamma_k \hat{\rho}_k(\theta)$ and 
using the convexity of the FI demonstrated above, we have 
\be
I_Q\big[\hat \rho(\theta) \big] = I\big [\hat \rho(\theta), \{ \hat E_{{\rm opt}, \hat \rho(\theta)}(\vect{\varepsilon}) \} \big] \leq \sum_k \gamma_k 
I\big [\hat \rho_k(\theta), \{ \hat E_{{\rm opt},\hat \rho(\theta)}(\vect{\varepsilon}) \}  \big]. \nonumber
\ee
Using again the definition of QFI, we have 
\be
I\big [\hat \rho_k(\theta), \{ \hat E_{{\rm opt},\hat \rho(\theta)}(\vect{\varepsilon}) \}  \big] \leq \max_{\hat E(\vect{\varepsilon})} 
I\big [\hat \rho_k(\theta), \{ \hat E (\vect{\varepsilon}) \}  \big] = I_Q\big[\hat \rho_k(\theta) \big], \nonumber
\ee
where $I_Q\big[\hat \rho_k(\theta) \big]= I\big [\hat \rho_k(\theta), \{ \hat E_{{\rm opt}, \hat \rho_k(\theta)}(\vect{\varepsilon}) \}  \big]$ and 
$\{ \hat E_{{\rm opt}, \hat \rho_k(\theta)}(\vect{\varepsilon}) \}$ is the optimal POVM for the state $\hat \rho_k(\theta)$.
Putting all these inequalities together, we recover Eq.~(\ref{QFIconvexity}). A direct but more involved proof can be found in \cite{FujiwaraPRA2001}.
}:
\be \label{QFIconvexity}
I_Q\bigg[ \sum_k \gamma_k \hat{\rho}_k(\theta) \bigg] \leq \sum_k \gamma_k I_Q\big[\hat \rho_k(\theta) \big].
\ee
This reflects the fact that mixing quantum states cannot increase the achievable
estimation sensitivity.

\subsubsection{Additivity}
\label{additivityf}

-- For $m$ independent subsystems and independent measurements
({\it i.e.} $m$ uncorrelated events) the FI is additive
\be \label{fisher_ind}
I(\theta) = \sum_{i=1}^m F_i(\theta),
\ee
where $F_i(\theta) = \sum_{\varepsilon_i}  \frac{1}{P_i(\varepsilon_i | \theta)}~\big(\frac{\partial P_i(\varepsilon_i | \theta)}{\partial \theta} \big)^2$
is the FI for the $i$th subsystem with the sum extending over all possible measurement results $\varepsilon_i$ ~\footnote{
Here we prove Eq.~(\ref{fisher_ind}). Using Eq.~(\ref{likfuninde}), we have
\beq 
I(\theta)= 
\sum_{\vect{\varepsilon}} P_1(\varepsilon_1|\theta) ... P_m(\varepsilon_m|\theta)
\sum_{i,j=1}^m \frac{\partial}{\partial \theta} \ln P_i(\varepsilon_i|\theta) 
\frac{\partial}{\partial \theta} \ln P_j(\varepsilon_j|\theta), \nonumber
\eeq
where the sum over $\vect{\varepsilon}$ runs over all possible values of $\varepsilon_1, ..., \varepsilon_m$.
We separate the sum over $i,j$ in the contribution $i=j$,
\beq
\sum_{i=1}^m \sum_{\vect{\varepsilon}} P_1(\varepsilon_1|\theta) ... P_m(\varepsilon_m|\theta) \Big( \frac{\partial}{\partial \theta} \ln P_i(\varepsilon_i|\theta) \Big)^2 = 
\sum_{i=1}^m \sum_{\varepsilon_i} \frac{1}{P_i(\varepsilon_i|\theta)} \Big( \frac{\partial P_i(\varepsilon_i|\theta)}{\partial \theta} \Big)^2 =
\sum_{i=1}^{m} F_i(\theta),  \nonumber
\eeq
and $i\neq j$,
\beq
&& \sum_{i,j=1; i\neq j}^m  \sum_{\vect{\varepsilon}} P_1(\varepsilon_1|\theta) ... P_m(\varepsilon_m|\theta)
\frac{\partial}{\partial \theta} \ln P_i(\varepsilon_i|\theta) 
\frac{\partial}{\partial \theta} \ln P_j(\varepsilon_j|\theta) = \nonumber \\
&& \quad\quad\quad\quad\quad\quad\quad\quad\quad\quad\quad\quad\quad\quad\quad\quad\quad\quad 
= \sum_{i,j=1; i\neq j}^m 
\bigg( \sum_{\varepsilon_i} \frac{\partial P_i(\varepsilon_i|\theta)}{\partial \theta} \bigg)
\bigg( \sum_{\varepsilon_j} \frac{\partial P_i(\varepsilon_j|\theta)}{\partial \theta} \bigg)=0, \nonumber
\eeq
which vanishes due to the normalisation property of the probability, see footnote $(^2)$.
}.
In the case of identical subsystems and identical POVMs, we have
\be
I(\theta) = m F(\theta),
\ee
with 
\be \label{FisherInformation}
F(\theta) \equiv \sum_{\varepsilon}  \frac{1}{P(\varepsilon | \theta)}~\Big(\frac{\partial P(\varepsilon | \theta)}{\partial \theta} \Big)^2.
\ee
The CRLB Eq. (\ref{Eq:crlb}) can thus be written as
 \begin{equation}
\label{crlb_5}
\big( \Delta \Theta_{\rm CR} \big)^2_\theta =
\frac{ \big( \frac{\partial \langle \Theta \rangle_\theta}{ \partial \theta}\big)^2 }
{m F(\theta)}.
\end{equation}
In the remaining of this Review we will mainly consider the case of independent identical observables and probes 
and will refer to $F(\theta)$ as the Fisher information.

The QFI is additive as well. For $m$ independent subsystems we have
\be \label{QFIadditivity}
I_Q\big[ \hat \rho^{(1)}(\theta)\otimes \hat \rho^{(2)}(\theta)\otimes ... \otimes \hat \rho^{(m)}(\theta) \big] = \sum_{i=1}^{m} I_Q[ \hat \rho^{(i)}(\theta) ],
\ee
where $I_Q[ \hat \rho^{(i)}(\theta) ] = \tr\big[ \hat{\rho}^{(i)}(\theta) \big( \hat L^{(i)}_{\theta} \big)^2 \big]$~\footnote{
Let us demonstrate Eq.~(\ref{QFIadditivity}). Taking the derivative of $\hat \rho(\theta)=\hat \rho^{(1)}(\theta)\otimes \hat \rho^{(2)}(\theta)\otimes ... \otimes \hat \rho^{(m)}(\theta)$
and expressing it in terms of the SLD for each subsystem, we have
\beq
\frac{\partial \hat \rho(\theta)}{\partial \theta} &=& \frac{\partial \hat \rho^{(1)}(\theta)}{\partial \theta}\otimes \hat \rho^{(2)}(\theta) ... \otimes \hat \rho^{(m)} (\theta)
+  \hat \rho^{(1)} (\theta) \otimes \frac{\partial \hat \rho^{(2)}(\theta)}{\partial \theta}\otimes ... \otimes \hat \rho^{(m)} (\theta) + ... \nonumber \\
&=& \frac{ \hat \rho(\theta) \big( \sum_{i=1}^m \hat L^{(i)}_\theta \big) +   \big( \sum_{i=1}^m \hat L^{(i)}_\theta \big) \hat \rho(\theta) }{2}, \nonumber
\eeq 
where $2\partial_\theta \hat{\rho}^{(i)}(\theta) = \hat{\rho}^{(i)}(\theta) \hat{L}^{(i)}_\theta + \hat{L}^{(i)}_\theta \hat{\rho}^{(i)}(\theta)$.
Taking into account Eq. (\ref{SLDdefinition}), we arrive at the equation  
$\hat \rho(\theta) \hat A  +   \hat A \hat \rho(\theta)=0$, where $\hat A \equiv \sum_{i=1}^N \hat L^{(i)}_\theta - \hat L_\theta$.
We now consider the complete eigenbasis $\{ \vert k \rangle\}$, $\sum_k \vert k \rangle \langle k \vert = \Eins$, 
such that $\hat \rho(\theta) = \sum_k p_k \vert k \rangle \langle k \vert$ with $p_k\geq 0$. 
We have  $(p_k + p_{k'}) A_{k, k'}=0$, where $A_{k, k'} = \langle k \vert \hat A \vert k' \rangle$.
According to this condition, $A_{k, k'}$ can be nonzero only if $p_k+p_{k'}=0$.
Remarkably, the matrix elements $\langle k \vert \hat L_{\theta} \vert k' \rangle$ for which $p_k+p_{k'}=0$ 
do not contribute to the QFI (see Sec.~\ref{QFIgeneral} and Eq.~(\ref{FQImatrixel}), in particular).
We thus conclude that, for our porpoises,  $\hat L_{\theta} = \sum_{i=1}^m \hat{L}^{(i)}_\theta$ and Eq.~(\ref{QFIadditivity}) is obtained by taking into account that $\tr[\hat{\rho}^{(i)}(\theta) \hat{L}^{(i)}_\theta]=0$,~$\forall i$.
}.
The QFI is saturated by the separable POVM Eq.~(\ref{indepmeas}) with elements 
$\hat{E}^{(i)}(\varepsilon)$ satisfying the conditions
$\Im \big( \tr[ \hat{\rho}^{(i)}(\theta) \hat L^{(i)}_{\theta} \hat{E}^{(i)}(\varepsilon)] \big) = 0$ 
and 
$\hat{\rho}^{(i)}(\theta) \big( \Eins - \lambda^{(i)}_{\theta, \varepsilon} \hat L^{(i)}_{\theta} \big) \hat{E}^{(i)}(\varepsilon) = 0$, 
$\forall \varepsilon_i$.
In the case of identical subsystems [$\hat{\rho}^{(i)}(\theta) = \hat{\rho}(\theta)$, $\forall\,i$] we have 
\be
I_Q\big[ \hat \rho^{(1)}(\theta)\otimes \hat \rho^{(2)}(\theta)\otimes ... \otimes \hat \rho^{(m)}(\theta) \big]= m F_Q\big[ \hat \rho (\theta) \big],
\ee
where 
\be \label{QFISI}
F_Q\big[ \hat \rho (\theta) \big] = \tr\big[\hat{\rho}(\theta) \hat L^2_{\theta} \big] \geq F(\theta) 
\ee
is the QFI for the single quantum state.
Finally, we recover the familiar form of the quantum Cram\'er-Rao bound
\be \label{QCRF}
(\Delta \Theta_{\rm QCR})^2_\theta = 
\frac{ \big( \frac{\partial \langle \Theta \rangle_\theta}{ \partial \theta}\big)^2 }
{m F_Q[ \hat \rho (\theta) ]}.
\ee
 

\subsubsection{The quantum Fisher information for mixed and pure states}
\label{QFIgeneral}

-- Here we determine the QFI and the SLD in terms of the 
the complete basis $\{\ket{k}\}$ such that 
$\hat{\rho}(\theta) = \sum_k p_k \ket{k}\bra{k}$,  with $p_k \geq 0$ and $\sum_k p_k=1$ \footnote{To simplify the notation 
we do not explicitly indicate the dependence of $\vert k \rangle$ and $p_k$ on $\theta$.}.
First, let us notice that, in this basis, Eq.~(\ref{QFISI}) can be written as 
\be \label{FQImatrixel}
F_Q\big[ \hat \rho (\theta)\big] = \sum_{k,k'} p_k \big\vert \langle k \vert \hat L_{\theta} \vert k' \rangle \big\vert^2 
= \sum_{k,k'} \frac{p_k+p_{k'}}{2} \big\vert \langle k \vert \hat L_{\theta} \vert k' \rangle \big\vert^2. 
\ee
Therefore, to calculate the QFI, it is sufficient to know the matrix elements $\langle k \vert \hat L_{\theta} \vert k' \rangle$
for the vectors $\vert k \rangle$ and $\vert k' \rangle$ such that $p_k + p_{k'} > 0$.
These matrix elements can be found using the definition of SLD, Eq.~(\ref{SLDdefinition}), giving 
$ \langle k \vert \hat L_{\theta} \vert k' \rangle = 2 \bra{k} \partial_\theta  \rho (\theta) \ket{k'} /(p_k+p_{k'})$.
From Eq.~(\ref{FQImatrixel}), we thus obtain \cite{HubnerPLA1992}
\be
F_Q\big[ \hat \rho (\theta) \big]
= 
\sum_{k,k'}
\frac{2}{p_k + p_{k'}} \big|  \bra{k} \partial_\theta  \rho (\theta) \ket{k'} \big|^2,
\ee
where the sum includes only terms for which $p_k + p_{k'} >0$.
In order to progress further we use
\begin{equation}
\partial_{\theta} \hat \rho(\theta) = 
\sum_k \big(\partial_{\theta} p_k \big) \proj{k}
+ \sum_k p_k \ket{\partial_\theta k} \bra{k}
+ \sum_k p_k \ket{k} \bra{\partial_\theta k},
\ee
where $\ket{\partial_\theta k} \equiv \partial_\theta \ket{k}$.
We thus have 
\begin{equation}
\bra{k} \partial_\theta  \rho (\theta) \ket{k'} = 
(\partial_{\theta} p_k) \delta_{k,k'} +
(p_k - p_{k'}) \bra{\partial_\theta k} k' \rangle,
\ee
where we have used 
$\partial_\theta \bra{k}k'\rangle = \bra{\partial_\theta k}k'\rangle + \bra{k}\partial_\theta k'\rangle =0$.
The SLD and the QFI become 
\be \label{SLDms}
\hat L_{\theta} =
\sum_k \frac{\partial_\theta p_k}{p_k} \proj{k}+
2 \sum_{k,k'} \frac{p_k - p_{k'}}{p_k + p_{k'}} \,
\ket{k}
\bra{\partial_\theta k} k' \rangle
\bra{k'},
\ee
and
\be \label{QFIms}
F_Q\big[ \hat \rho (\theta) \big]
= \sum_k \frac{(\partial_\theta p_k)^2}{p_k}
+ 2 \sum_{k,k'} \frac{(p_k - p_{k'})^2}{p_k + p_{k'}}
\big\vert \bra{\partial_\theta k} k' \rangle \big\vert^2,
\ee
respectively \cite{BraunsteinPRL1994}.
These equations simplify for pure states  $\vert \psi(\theta) \rangle$.  
We can write $\Eins=\vert \psi (\theta) \rangle \langle \psi (\theta)\vert + \sum_{k_\perp} \vert k_\perp \rangle \langle k_{\perp} \vert$, 
where $\{ \vert k_\perp \rangle \}$ is a basis of the Hilbert space orthogonal to the vector $\vert \psi(\theta) \rangle$, and thus 
formally $\hat\rho(\theta)=\vert \psi(\theta) \rangle \langle \psi(\theta) \vert + \sum_{k_\perp} p_{k_\perp} \vert k_\perp \rangle \langle k_{\perp} \vert$, 
where $p_{k_\perp}=0$ $\forall {k_\perp}$.
Equations~(\ref{SLDms}) and~(\ref{QFIms}) give 
\be \label{SLDms2}
\hat L_{\theta} = 2\ket{\psi}\bra{\partial_\theta \psi} + 2\ket{\partial_\theta \psi}\bra{\psi},
\ee
and 
\be \label{QFIms2}
F_Q\big[ \vert \psi (\theta) \rangle \big]= 4 \big( \bra{\partial_\theta \psi}\partial_\theta \psi\rangle - \vert \bra{\partial_\theta \psi}\psi \rangle \vert^2 \big),
\ee
respectively. Notice that we can obtain the same expressions 
also using the relation $\hat \rho^2(\theta) = \hat \rho(\theta)$, valid for pure states. 
The derivative with respect to the parameter is
$\partial_\theta \hat \rho = \partial_\theta \hat \rho^2= (\partial_\theta \hat \rho) \hat \rho +  \hat \rho (\partial_\theta \hat \rho)$.
We can identify $\hat L_{\theta} = 2\partial_\theta \hat \rho$, which coincides with Eq.~(\ref{SLDms2}), 
and obtain Eq.~(\ref{QFIms2}) by directly applying the definition (\ref{QFISI})
and using the relation $\partial_\theta \langle \psi \vert \psi \rangle = \langle \partial_\theta \psi \vert \psi \rangle + \langle \psi \vert \partial_\theta \psi \rangle = 0$.

It is interesting to point out that the QFI for pure states, Eq.~(\ref{QFIms2}), can be saturated, in the limit $\theta \to 0$, 
by projective measurements on the probe state $\vert \psi_0 \rangle \equiv \vert \psi(0) \rangle$ and on the orthogonal subspace.
We can easily see this explicitly. Let us consider the POVM set $\{ \hat{\Pi}_k \}$ with $\hat{\Pi}_1 = \vert \psi_0 \rangle \langle \psi_0 \vert$,
and  $\hat{\Pi}_k \vert \psi_0 \rangle=0$ for $k\neq 1$. We have 
\be
F(\theta) =
4 \sum_k 
\frac{
\big( {\rm Re}[ \langle \partial_\theta \psi(\theta) \vert \hat{\Pi}_k \vert \psi(\theta) \rangle] \big)^2 
}{ \langle \psi(\theta) \vert \hat{\Pi}_k \vert \psi(\theta) \rangle }.
\ee
We now calculate the limit $\theta \to 0$. For $\theta=0$, the terms $k = 1$ in the above 
sum is equal to zero and does not contribute to the FI, while the terms $k \neq 1$ are undetermined (0/0). 
We thus evaluate the limit using de l'H\^opital's rule:
\be
\lim_{\theta \to 0} F(\theta) = 
4 \lim_{\theta \to 0} \sum_{k\neq 1} 
\frac{
\big( {\rm Re}[ \langle \partial_\theta \psi(\theta) \vert \hat{\Pi}_k \vert \psi(\theta) \rangle] \big)^2 
}{ \langle \psi(\theta) \vert \hat{\Pi}_k \vert \psi(\theta) \rangle }
 =  4 \sum_{k\neq1} \langle \partial_\theta \psi_0 \vert \hat{\Pi}_k \vert \partial_\theta \psi_0 \rangle = F_Q\big[ \vert \psi_0 \rangle \big],
\ee
where we have used $\sum_{k\neq 1} \hat{\Pi}_k = \Eins -  \vert \psi_0 \rangle \langle \psi_0 \vert$ and Eq.~(\ref{QFIms2}).
The saturation of the QFI requires the POVM set to consist of a minimum of two elements 
($\vert \psi_0 \rangle\langle \psi_0\vert $ and $\Eins- \vert \psi_0 \rangle \langle \psi_0 \vert $).
This result is due to the strong contribution to the Fisher information of outcomes with low probabilities
(we remind that $\langle \psi(\theta) \vert \hat{\Pi}_k \vert \psi(\theta) \rangle=0$ for $k\neq 1$), due to the fast change of these probabilities with respect to the parameter. 


\subsubsection{The quantum Fisher information for unitary transformations}
\label{unitaryQFI}

-- We consider now the relevant case of unitary transformations
\begin{equation} \label{rhounit}
\hat \rho(\theta) = e^{-i \theta \hat H} \hat \rho_0 e^{+i \theta \hat H},
\ee
where $\hat H$ is a Hermitian operator and
$\hat \rho_0$ is the ``probe'' or the ``input'' state.
The unitary transformation $e^{-i \theta \hat H}$ describes quantum mechanically our interferometer.
As a first important simplification, we note that, upon diagonalizing 
the probe state $\hat \rho_0 = \sum_{k} p_k \ket{k}\bra{k}$, we obtain 
$\hat \rho(\theta) = \sum_k p_k e^{-i \theta \hat H} \proj{k} e^{+i \theta \hat H}$:
the unitary evolution do not change the eigenvalues $p_k$.
The SLD is 
given by $\hat L_\theta = e^{-i \theta \hat H} \hat L_0 e^{i \theta \hat H}$,
where, according to Eq.~(\ref{SLDdefinition}), $\hat L_0$ satisfies the 
equation 
\be 
\{ \hat \rho_0, \hat L_0 \} = 2i[\hat\rho_0, \hat H].
\ee
The QFI, expressed in terms of $\hat L_0$, 
\begin{equation}
F_Q\big[ \hat \rho_0, \hat H \big] = (\Delta \hat L_0)^2,
\ee
does not depend explicitly on $\theta$.
The dependence on $\theta$ can still be present in the POVM that saturates the QFI.
Equations~(\ref{SLDms}) and (\ref{QFIms}) give
\begin{equation}
\hat L_0 =
2i \sum_{k,k'} \frac{p_k - p_{k'}}{p_k + p_{k'}} \,
\ket{k}\bra{k}\hat H \ket{k'}
\bra{k'},
\ee
and
\begin{equation} \label{QFImsUnit}
F_Q\big[ \hat \rho_0, \hat H \big] = 2 \sum_{k,k'} \frac{(p_k - p_{k'})^2}{p_k + p_{k'}} 
\big\vert \bra{k} \hat H \ket{k'} \big\vert^2,
\ee
where the sum extends to $p_k + p_{k'}\neq 0$.
For pure states $\hat \rho_0 = \proj{\psi_0}$ these equations further simplify to [see Eqs.~(\ref{SLDms2}) and (\ref{QFIms2})]
\begin{equation}
\hat L_0 = 2 i \proj{\psi_0} \hat H - 2i \hat H \proj{\psi_0},
\ee
and 
\begin{equation} \label{QFIpsUnit}
F_Q\big[ \vert \psi_0 \rangle, \hat H \big] = 4 (\Delta \hat H)^2.
\ee
We want to further investigate the relation between Eq.~(\ref{QFImsUnit})
and Eq.~(\ref{QFIpsUnit}). 
We already know, from Sec.~(\ref{convexityf}) that the FI
is convex and thus reaches its maximum value on pure states.
We thus expect $F_Q[\hat \rho_0, \hat H] \leq 4 (\Delta \hat H)^2$,
with equality for pure states. 
It is interesting to recover explicitly this result.
First, notice that, since the numerator 
in Eq.~(\ref{QFImsUnit}) contains the difference $p_k -p_{k'}$,
we can replace $\hat H \to \hat H - h$, with $h$ an arbitrary complex number,
without changing the value of the QFI.
Since $p_k$ are non negative numbers and $p_k+p_{k'}\neq 0$, we have 
$\frac{(p_k - p_{k'})^2}{p_k + p_{k'}} \leq p_{k} + p_{k'}$.
Therefore, from Eq.~(\ref{QFImsUnit}) we have
\be \label{FQPSbound}
F_Q\big[ \hat \rho_0, \hat H \big] \leq 4 \sum_{k,k'} p_k \big\vert \bra{k} \hat H - h \ket{k'} \big\vert^2 = 
4 \big( \Delta \hat H \big)^2 + 4 \big\vert  \langle \hat H \rangle - h  \big\vert^2.
\ee
The right hand side of Eq.~(\ref{FQPSbound}) is minimised for $h=\langle \hat H \rangle$, giving $F_Q\big[ \hat \rho_0, \hat H \big] \leq 4 \big( \Delta \hat H \big)^2$, with equality for pure states. 
Summarizing, for unitary evolution, we find the following chain of inequalities:
\begin{eqnarray} \label{QFchain}
F(\theta) \leq F_Q\big[ \hat \rho_0, \hat H \big] = \tr\big[ \rho_0 \hat{L}_0^2 \big] \leq 4 \big( \Delta \hat{H} \big)^2.
\end{eqnarray}


\subsubsection{Lower bounds}
\label{bounds}

-- Let us consider an arbitrary diagonal operator $\hat{M} = \sum_\mu c_\mu~|\mu \rangle \langle \mu | $, where $\{|\mu \rangle\}$
is a complete orthonormal basis and $\{c_\mu\}$ complex numbers. 
The FI is bounded by: 
\begin{equation} \label{MomentIneq}
F(\theta) \geq 
\frac{\big| \frac{\ud \langle \hat M \rangle}{\ud \theta} \big|^2}
{\sum_\mu | c_\mu - f(\theta) |^2 \, P(\mu | \theta)},
\end{equation}
where $\langle \hat M \rangle=  \tr[ \hat{M} \hat{\rho}(\theta)]  = \sum_\mu c_\mu \, P(\mu | \theta)$ and 
$f(\theta)$ is an arbitrary function of $\theta$
\footnote{
Here we prove Eq.~(\ref{MomentIneq}).
We have $\frac{\ud \langle \hat M \rangle}{\ud \theta} = 
\sum_\mu c_\mu \frac{\ud P(\mu \vert \theta)}{\ud \theta} =
\sum_\mu ( c_\mu - f(\theta)) \frac{\ud P(\mu \vert \theta)}{\ud \theta}$,
where $P(\mu \vert \theta) = \langle \mu \vert  \rho(\theta) \vert \mu \rangle$,
$f(\theta)$ is an arbitrary function of $\theta$ and 
we have used the property $\sum_\mu f(\theta) \frac{\ud P(\mu \vert \theta)}{\ud \theta}=
f(\theta) \frac{\ud }{\ud \theta} \sum_\mu P(\mu \vert \theta)=0$.
We now apply the Cauchy-Schwarz inequality 
$|\sum_\mu a_\mu b_\mu|^2 \leq (\sum_\mu \vert a_\mu^2|) (\sum_\mu \vert b_\mu^2|)$, 
with 
$a_\mu = \frac{1}{\sqrt{P(\mu \vert \theta)}} \frac{\ud P(\mu \vert \theta)}{\ud \theta}$ and 
$b_\mu = ( c_\mu - f(\theta)) \sqrt{P(\mu \vert \theta)}$.
This gives $|\frac{\ud \langle \hat M \rangle }{\ud \theta}|^2 \leq 
F(\theta) \, \sum_\mu | c_\mu - f(\theta)|^2 P(\mu \vert \theta)$ and we thus recover Eq.~(\ref{MomentIneq}). 
}.
If $\hat M$ is a Hermitian operator ($c_\mu$ are real) and 
the phase shift is provided by a unitary operator $e^{-i \hat{H} \theta}$, with $\hat{H}$ being the generator of the phase
shift, the Ehrenfest theorem gives $i \frac{\ud \langle \hat M \rangle}{\ud \theta} = \langle [\hat{M}, \hat{H}]\rangle$.
With the further choice $ f(\theta) = \langle \hat M \rangle$ we get:
\be \label{MomentIneq1}
F(\theta) \geq \frac{\big| \langle [ \hat{M},\hat{H}]\rangle \big|^2}
{(\Delta \hat M)^2}.
\ee
As a second example, if we choose $\hat M$  to be an arbitrary unitary operator $\hat U =  \sum_\mu e^{i \mu}~|\mu \rangle \langle \mu | $ 
and $f(\theta) = 0$ we have
\be \label{MomentIneq2}
F(\theta) \geq \bigg| \frac{\ud \langle \hat U \rangle}{\ud \theta} \bigg|^2 = \big| \langle [\hat{H}, \hat{U}]\rangle |^2.
\ee
where the second inequality again holds for unitary transformations. 


\subsection{The Maximum Likelihood Estimator}
\label{MaxLik}

In this section we discuss one of the most important estimators: the maximum likelihood (ML). 
It is defined as the phase value 
\be \label{MLdef}
\Theta_{\rm ML}( \vect{\varepsilon}) = 
\arg\Big[ \max_{\varphi} P(\vect{\varepsilon}\vert\varphi) \Big],
\ee
which maximizes the likelihood (as a function of the free variable $\varphi$) for a given sequence $\vect{\varepsilon}$ of $m$ measurements. 
The ML can be seeked as the solution of 
\be \label{LMLdef}
\frac{\partial P(\vect{\varepsilon}| \varphi) }{\partial \varphi} \bigg|_{\Theta_{\rm ML}}=0, \qquad \qquad
\frac{\partial^2 P(\vect{\varepsilon}| \varphi) }{\partial \varphi^2} \bigg|_{\Theta_{\rm ML}}<0.
\ee
It is equivalent, and often convenient, to calculate the ML by maximizing the log-likelihood 
$L\big( \vect{\varepsilon}| \varphi \big) \equiv \ln P(\vect{\varepsilon}| \varphi)$.
Since the measurement results $\vect{\varepsilon}$ 
are random outcomes distributed according to $P(\vect{\varepsilon}|\theta)$, also the 
values of $\Theta_{\rm ML}$ are randomly distributed.
In order to calculate the sensitivity of the ML estimator, 
we need to build up a histogram with the outcomes $\Theta_{\rm ML}$ obtained by repeating a large number of 
times the sequence of the $m$ measurements.
In the following we discuss the main asymptotic properties of the ML estimator \cite{LehmannBOOK1998} and, most importantly, show that 
the histogram would converge to a smooth Gaussian distribution with a width equal to the CRLB.

\subsubsection{Asymptotic Consistency}
\label{ConsML}

-- Here we demonstrate that the $\Theta_{\rm ML}(\vect{\varepsilon})$,
obtained with $m$ independent measurements (equally distributed), tends in probability to the true value $\theta$ 
of the parameter, as $m\to \infty$:
\be \label{ConsistencyML}
\lim_{m \to \infty} {\rm Pr} \Big[ \big\vert \Theta_{\rm ML}(\vect{\varepsilon}) - \theta \big\vert > \delta \Big]=0,
\ee
for any arbitrarily small $\delta$.
In other words, the ML estimator is {\it asymptotic consistent}. 
To prove it, note that, since $\Theta_{\rm ML}$ is defined as the maximum of $P(\vect{\varepsilon}|\varphi)$, it is also the maximum 
of $L(\vect{\varepsilon}|\varphi) = \sum_{i=1}^m \ln P(\varepsilon_i\vert \theta)$ divided by $m$ and subtracted by a constant:
\beq
\Theta_{\rm ML}( \vect{\varepsilon}) =
\arg\bigg[ \frac{1}{m} \max_{\varphi} \Big(
 \ln P(\vect{\varepsilon}|\varphi) - \ln P(\vect{\varepsilon} \vert \theta) \Big)\bigg] 
= \arg\bigg[ \min_{\varphi} \Big(
\frac{1}{m} \sum_{i=1}^m \ln \frac{P(\varepsilon_i\vert \theta)}{P(\varepsilon_i\vert \varphi)} \Big)\bigg]. \nonumber \\ \label{MLcons1}
\eeq
Taking the limit $m \to \infty$, the right hand side of Eq.~(\ref{MLcons1}) converges in probability, for the law of large 
numbers~\footnote{
Consider $m$ independent and identically distributed random variables
$x_1, ..., x_m$ with mean value $\mu$.
The {\it weak law of large numbers} states that the sample 
average converges in probability towards the expected value $\mu$,
$S_m = \frac{1}{m} \sum_{i=1}^m x_i \to \mu$ as $m\to \infty$. 
That is to say that for any positive number $\delta$, 
$\lim_{m \to \infty} {\rm Pr}[ \vert S_m - \mu \vert > \delta] = 0$,
i.e. the average $S_m$ will be found in the interval 
$[\mu-\delta, \mu+\delta]$ with unit probability, 
no matter how small $\delta$, provided that $m$ is sufficiently large.
The weak law of large numbers leaves open the possibility to have 
$|S_m -\mu| > \delta$, although the probability to have such 
situation is infrequent.
The {\it strong law of large numbers} states that the sample average converges 
almost surely (or strongly) to the expected value, 
${\rm Pr}[ \lim_{m \to \infty} S_m = \mu ] = 1$.
This is called the strong law because random variables which converge strongly (almost surely) are guaranteed to converge weakly (in probability).
In particular, the strong law implies, with unit probability, that
the inequality $|S_m -\mu| < \delta$ holds for any $\delta>0$ and for large enough $m$.
},
to $\arg[ \min_{\varphi} K( P_\theta || P_\varphi )]$, where 
\be \label{KLML}
K( P_\theta || P_\varphi ) = \sum_\varepsilon P(\varepsilon\vert \theta) \ln \frac{P(\varepsilon\vert \theta)}{P(\varepsilon\vert \varphi)}
\ee
is the Kullback-Leibler divergence.
As the logarithm is a strictly concave function,
we have~\footnote{
Given a real concave function $f(x)$ [{\it i.e.} such that $f(\lambda_1 x_1 + \lambda_2 x_2)
\geq \lambda_1 f( x_1) + \lambda_2 f(x_2)$, 
$\forall x_1, x_2$ and $\lambda_1+\lambda_2=1$], Jensen's inequality states that 
$f(\sum_i \lambda_i x_i/\sum_j \lambda_j)\geq \sum_i \lambda_i f(x_i)/ \sum_j \lambda_j$, where $\lambda_i$ are positive weights.
This inequality, with $f(x) = \ln(x)$, $x = P(\varepsilon\vert \varphi)=P(\varepsilon\vert \theta)$ and
weights $P(\varepsilon\vert \theta)$, gives Eq. (\ref{KLMLineq}).
}, showing that 
\be \label{KLMLineq}
K( P_\theta || P_\varphi ) 
\geq  0, 
\ee
with equality if and only if $P(\varepsilon\vert \varphi)=P(\varepsilon\vert \theta)$ $\forall \varepsilon$, 
i.e. if and only if $\varphi=\theta$.
In conclusion, in the limit $m \to \infty$, $\Theta_{\rm ML} = \arg[ \min_{\varphi} K( P_\theta || P_\varphi )] = \theta$.

\subsubsection{Asymptotic normality and efficiency}
\label{MLnormality}

-- The key role played by the ML in parameter estimation is due to its asymptotic properties for independent measurements, first discussed by  
Fisher \cite{Fisher1922}. For sufficiently large $m$,
the distribution of the ML estimator tends to a
Gaussian centered to the true value $\theta$ of the phase shift and of
variance equal to the inverse FI:
\begin{equation} \label{PML}
P(\Theta_{\rm ML} | \theta) \to \sqrt{\frac{m F(\theta)}{2 \pi}}  
e^{- \frac{m F(\theta)}{2} (\Theta_{\rm ML} - \theta)^2}, \quad\quad {\rm for}\,\,m \to \infty.
\end{equation}
The ML is thus {\it asymptotically efficient}: in the limit $m \to \infty$ 
it saturates the CRLB.
To prove this, let us expand $\partial L\big( \vect{\varepsilon}| \varphi \big)/\partial \varphi$ 
in Taylor series around $\theta$:
\be \label{ML_Taylor}
\frac{\partial L\big(\vect{\varepsilon}|\varphi\big)}{\partial \varphi} 
=
\frac{\partial L\big(\vect{\varepsilon}|\varphi\big)}{\partial \varphi} \bigg|_{\theta}
+
\frac{\partial^2 L\big(\vect{\varepsilon}|\varphi\big)}{\partial \varphi^2} \bigg|_{\theta}
\big(\varphi - \theta \big) + O\big(\varphi - \theta \big)^2.
\ee
The asymptotic consistency of the ML guarantees that, for $m$ sufficiently large, 
$\Theta_{\rm ML}$ is sufficiently close to $\theta$ so to neglect higher orders in the expansion. 
We now evaluate Eq.~(\ref{ML_Taylor}) at $\varphi=\Theta_{\rm ML}$. 
Taking into account Eq.~(\ref{LMLdef}), we find 
\be
\frac{\partial^2 L\big(\vect{\varepsilon}|\varphi\big)}{\partial \varphi^2} \bigg|_{\theta}
\big(\Theta_{\rm ML} - \theta \big) = - \frac{\partial L\big(\vect{\varepsilon}|\varphi\big)}{\partial \varphi} \bigg|_{\theta},
\label{expa}
\ee 
up to the leading order in the expansion.
Equation~(\ref{expa}) can be written in terms of the single-measurement likelihood functions  
$\ell(\varepsilon_i \vert \varphi ) \equiv \ln P(\varepsilon_i \vert \varphi)$ [we assume here that the $m$ measurements are identically distributed and recall that $L\big( \vect{\varepsilon}| \varphi \big)= \sum_{i=1}^m \ell\big(\varepsilon_i \vert \varphi \big)$]:
\be \label{MLproof1}
\sum_{i=1}^m \frac{\partial^2 \ell \big(\varepsilon_i|\varphi\big)}{\partial \varphi^2} \bigg|_{\theta}
\big(\Theta_{\rm ML} - \theta \big) = 
- \sum_{i=1}^m \frac{\partial \ell\big(\varepsilon_i|\varphi\big)}{\partial \varphi} \bigg|_{\theta}.
\ee
Let us now introduce  
\be
S_m \equiv - \frac{1}{m} \sum_{i=1}^m \frac{\partial^2 \ell \big(\varepsilon_i|\varphi\big)}{\partial \varphi^2} \bigg|_{\theta}
\big(\Theta_{\rm ML} - \theta \big).
\ee
In the limit $m\to \infty$, the law of large numbers [see footnote ($^{13}$)] tells us that 
$\frac{1}{m} \sum_{i=1}^m \frac{\partial^2 \ell (\varepsilon_i|\varphi )}{\partial \varphi^2} \big\vert_{\theta}$
converges to $ \big\langle  
\frac{\partial^2 \ell (\varepsilon|\varphi )}{\partial \varphi^2} \big\vert_{\theta} \big\rangle
= - F(\theta)$, so that 
\be \label{MLproof6}
S_m = F(\theta)(\Theta_{\rm ML} - \theta), \quad \quad {\rm for}\,\, m\to \infty.
\ee
According to Eq.~(\ref{MLproof1}), 
$S_m $ is the sample average 
of the random variable $\partial \ell(\varepsilon_i|\varphi)/\partial \varphi\vert_\theta$, therefore
the central limit theorem~\footnote{
Consider $m$ independent and identically distributed random variables
$x_1, ..., x_m$ with $S_m \equiv \frac{1}{m} \sum_{i=1}^m x_i$ and finite mean value $\mu$ and variance $\sigma^2$. 
The {\it central limit theorem} states that the random variables $\sqrt{n}(S_m - \mu)$ converge in distribution to 
the normal distribution of zero mean and variance $\sigma^2$. 
in other words, $S_m$ behaves as a random variable with normal distribution, mean $\mu$ and variance $\sigma^2/m$:
$P( S_m ) = \frac{1}{\sqrt{2\pi\sigma^2/m}} e^{-\frac{m(S_m - \mu)^2}{2 \sigma^2}}$.} 
guarantees that $S_m$ becomes normally distributed,
\be \label{MLproof5}
P\big(S_m\big) = \frac{1}{\sqrt{2 \pi \sigma^2/m}} e^{-\frac{m (S_m-\mu)^2}{2 \sigma^2} }, \quad \quad {\rm for}\,\, m\to \infty,
\ee
where the mean value $\mu$ and the variance $\sigma^2$ are 
\beq \label{MLproof4}
\mu = \sum_{\varepsilon}  
\frac{\partial P(\varepsilon\vert \varphi)}{\partial \varphi} \bigg\vert_{\theta} = 0, 
\quad \quad
\sigma^2=\sum_{\varepsilon} P(\varepsilon\vert \theta) \bigg( \frac{1}{P(\varepsilon\vert \theta)} 
\frac{\partial P(\varepsilon\vert \varphi)}{\partial \varphi} \bigg\vert_{\theta}\bigg)^2 = 
F(\theta). 
\eeq
Combining Eqs.~(\ref{MLproof6})-(\ref{MLproof4}), we arrive at Eq.~(\ref{PML}).
Finally, since $\vert \Theta_{\rm ML} - \theta \vert \propto 1/m$, this result is consistent with the Taylor expansion~(\ref{ML_Taylor}),
which neglects higher order terms of $\vert \Theta_{\rm ML} - \theta \vert $, in the limit $m \to \infty$.

To summarize, Eqs.~(\ref{ConsistencyML}) and~(\ref{PML}) show that the ML estimator is asymptotically unbiased 
(tends to the true value of the phase shift) and efficient
(saturates the CRLB). 
Notice however that the ML estimator can be biased not only for small $m$, but also for {\it any finite} value of $m$. 
Yet, in the limit $m \to \infty$, the ML is as good or better than any other estimator. 
When the number of measurements $m$ is small (so to be outside the central limit condition) or the ML is 
biased for any finite value of $m$, it is possible to perform a phase estimation with a Bayesian approach (see Sec.~\ref{Bay}).
We finally point out that in the context of phase estimation the maximum likelihood analysis has been 
used in several experiments \cite{KacprowiczNATPHOT2010, KrischekPRL2010, HradilPRL1996, ZawiskyJPA1998, PezzePRL2007}.


\subsection{The Method of Moments}
\label{MethodMom}

Performing a ML phase estimation requires the knowledge of the probability $P(\varepsilon|\theta)$ 
for any $\theta$ and for all possible measurement result $\varepsilon$. 
From a practical point of view, these probabilities can 
be provided by the theory taking into account the experimental noise or directly retrieved
by a calibration of the interferometer. 
However, the extraction of the full $P(\varepsilon|\theta)$ can in some cases be difficult. 
How can we build an efficient phase estimation protocol if we have access to a limited 
amount of information about the system ?
Here we consider the measurement of an observable, say $\hat{M}$, with $\theta$-dependent 
mean value $\langle \hat{M} \rangle_\theta$ and variance $(\Delta \hat{M})^2_\theta$.
The observable $\hat{M}$ can be, for instance, the relative number of 
particles at the output ports of a Mach-Zehnder interferometer.
Let us consider $m$ measurements of $\hat{M}$, with results $\mu_1,\mu_2, \ldots ,\mu_m$,
and take the mean value
\begin{eqnarray} \label{mm}
M_m \equiv {1 \over m} \sum_{i=1}^m \mu_i.
\end{eqnarray}
By applying the central limit theorem [see footnote $(^{15})$], we find that
the probability distribution of $M_m$ is 
\begin{eqnarray} \label{Sm} 
P\big(M_m \vert \theta\big)  = 
\sqrt{\frac{m}{2 \pi (\Delta \hat{M}_\theta)^2}}
e^{ -\frac{ m (M_m-\langle \hat{M}\rangle_\theta)^2}{ 2 (\Delta \hat{M})_\theta^2 }} , \qquad {\rm for}\,\, m\to \infty.
\end{eqnarray}  
The very powerful result here is that, even if we do not know the full probability distribution $P(\mu|\theta)$, the central limit provides the 
probability for the stochastic variable $M_m$ in terms of its mean and variance. 
We can then apply all the machinery developed in the previous sections to choose a good estimator and determine its sensitivity. 
It should be obvious by now that, asymptotically in $m$, an excellent estimator is the maximum (we call it $\Theta$ here) 
of the likelihood function (\ref{Sm}) with $M_m$ as stochastic variable:
\be \label{Mom_ML}
\frac{\partial \ln P(M_m \vert \varphi)}{\partial \varphi}\Big\vert_{\Theta} =0, \qquad 
\frac{\partial^2 \ln P(M_m \vert \varphi)}{\partial \varphi^2}\Big\vert_{\Theta} < 0.
\ee
The left hand side of Eq.~(\ref{Mom_ML}) provides
\be
\big(M_m-\langle \hat{M}\rangle_\theta\big) 
\bigg( \frac{\partial \langle \hat{M} \rangle_\theta}{\partial \theta} + 
\frac{M_m-\langle \hat{M}\rangle_\theta}{(\Delta \hat{M})_\theta} \frac{\partial (\Delta \hat{M})_\theta}{\partial \theta}
 \bigg) =0,
\ee
to leading order in $m$.
This equation has two solutions but, among them, only 
\be
M_m = \langle \hat{M}\rangle_\theta,
\ee
satisfies the right hand side of Eq.~(\ref{Mom_ML}).
Introducing the function $f(\varphi)\equiv \langle \hat{M}\rangle)_{\varphi}$,
the estimator $\Theta$ is thus the value of the parameter for which
$f(\Theta) = M_m$:
\beq \label{MomMLEst}
\Theta &=& f^{-1}(M_m). 
\eeq
Obviously, the inversion is possible only in the phase intervals where $f(\varphi)$ is monotone.
In the limit $m \to \infty$, we have that $M_m \to \langle \hat{M}\rangle_\theta$ and thus $\Theta \to \theta$.
Furthermore, in this limit, the sensitivity $\Delta \Theta$ is given by the CRLB~\footnote{
We can see this explicitly from Eq.~(\ref{crlb_saturation}), i.e. the necessary and sufficient condition for the saturation of the CRLB, 
which here reads 
$\frac{\partial \ln P( M_m \vert \theta)}{ \partial \theta} = 
\lambda(\theta) \, \big(\Theta- \langle \Theta \rangle_\theta \big)$, $\forall \, M_m$.
In the limit $m \to \infty$, $M_m \to \langle \hat M \rangle_\theta$ and $\Theta \to \theta$: both sides of this equation 
vanish.
} with FI 
\beq
F(\theta) &=& \int \ud M_m \, \frac{1}{P(M_m| \theta)} \bigg( \frac{\partial P(M_m | \theta)}{\partial \theta} \bigg)^2. 
\eeq
To the leading order in $m \to \infty$, we get 
$F(\theta) = \frac{m}{(\Delta \hat{M})^2_\theta} \big( \frac{\partial \langle \hat{M} \rangle_\varphi}{\partial \varphi} \big\vert_{\theta} \big)^2$,
and therefore 
\be \label{MomCRLB}
(\Delta \Theta)^2 = \frac{(\Delta \hat{M})^2_\theta}{m \big(\frac{\partial \langle \hat{M} \rangle_\varphi}{\partial \varphi} \big\vert_{\theta}  \big)^2}, \quad \quad 
{\rm for}\,\, m \gg 1.
\ee
To summarize, when only the first moments of the probability distribution are experimentally accessible or theoretically known, 
the best estimator in the central limit is Eq.~(\ref{MomMLEst})
with an expected sensitivity given by Eq.~(\ref{MomCRLB}).
Equation~(\ref{MomCRLB}) 
is widely used in the literature to calculate the phase sensitivity of an interferometer for various input states
\cite{CavesPRD1981,Yurke_1986,BollingerPRA1996,DowlingPRA1998,WinelandPRA1994,KimPRA1998,GerryPRA2003,PezzePRA2005}.
We have shown that this sensitivity saturates the CRLB and thus it is the best we can have when 
measuring the average moment $M_m$. 
It is interesting to see that this equation can be obtained heuristically from an error propagation~\footnote{
Let us expand $f(\Theta)$ in Taylor series around the true value of the phase shift:
$f(\Theta) = f(\theta)+ \frac{\partial f(\varphi)}{\partial \varphi} \vert_{\theta} (\Theta - \theta) + O(\Theta- \theta)^2$ or, 
equivalently, 
$M_m = \langle \hat{M}\rangle_\theta + \frac{\partial \langle \hat{M}\rangle_{\varphi} }{\partial \varphi}\big\vert_{\theta} (\Theta-\theta) + O(\Theta-\theta)^2$.
We can now identify $(M_m - \langle \hat{M}\rangle_\theta)^2 \approx (\Delta \hat M)^2_\theta/m$ and
$(\Theta-\theta)^2 \approx (\Delta \Theta)^2$. We thus recover Eq.~(\ref{MomCRLB}).
}. 
Of course, as expected on a general ground and proven by Eq.~(\ref{MomentIneq}),  the estimator
Eq.~(\ref{MomMLEst}) is optimal when we have only access to the average moment $M_m$. 
Nevertheless it might not be optimal if we could retrieve the probability distribution of the single measurement results \cite{PezzePRL2007}.


\section{Bayesian phase estimation}
\label{Bay}

In this section we discuss the Bayesian approach to estimate an unknown (but fixed) value $\theta$ of the phase shift.
Bayesian estimation is based on an interpretation of probability
which is different from the standard frequentist view.
In the later case, the probability is defined as the infinite-sample 
limit of the outcome frequency of an observed event. 
In the Bayesian view, the probability is a normalized ``a posteriori'' functions of the parameter, 
obtained for a given measurement result.
The crucial point is that, in order to get this probability, the Bayesian framework introduces a prior
with the aim of quantifying our ``a priori'' (i.e. before making any measurement) knowledge about the true value of the parameter. 
From a foundational point of view, this means that the Bayesian setting introduces a subjective 
interpretation of the probability, defined as our ``a posteriori'' (i.e. after collecting experimental outcomes) measure of ignorance (or knowledge). 
In practice if the true value of the phase shift $\theta$ is unknown over the full $[0,2 \pi]$ interval, 
we typically consider a flat prior probability $P(\theta) = 1 / 2 \pi$ to express our maximum ignorance. 

An appealing property of Bayesian inference is the possibility to draw the uncertainty of an estimate 
from the specific sequence of observed data (and not from the reconstruction of
histograms,
as in the frequentist case, requiring the collection of a large number of data sequences).
In addition, the Bayesian method allows to eliminate nuisance parameters,
reducing the dimension of data analysis.
Remarkably, the Bayesian method is asymptotically consistent: 
as the number of measurements increases, the posterior probability distribution assign more weight 
in the vicinity of the true value.
It is possible to demonstrate that, in the central limit, the posterior probability 
becomes normally distributed, centred at the true value of the parameter and with 
a variance inversely proportional to FI. 

In the context of phase estimation, Bayesian inference has been 
used in optical \cite{XiangNATPHOT2010, KrischekPRL2010, PezzePRL2007} 
and neutron \cite{HradilPRL1996, ZawiskyJPA1998} experiments
and applied theoretically to calculate the phase sensitivity achievable with different quantum states \cite{PezzePRA2006, PezzeEPL2007, OlivaresJPB2009}, 
eventually making use of adaptive strategies \cite{BerryPRL2000}.

\subsection{Bayesian inference}
\label{BayestChapter:Inference}

The cornerstone of Bayesian inference \cite{GhoshBOOK} is the Bayes' theorem.
Let us consider $\varphi$ as a continuous {\it random} variable defined in the interval $[a,b]$
and $\vect{\varepsilon} = \{ \varepsilon_1, \varepsilon_2, \ldots, \varepsilon_m\}$ the result of 
$m$ measurements.
We denote with $P\big(\varphi , \vect{\varepsilon}\big)$ 
the joint probability density function of having a random parameter $\varphi$ 
{\it and} a random outcome $\vect{\varepsilon}$. 
The joint probability density is symmetric, $P\big(\varphi , \vect{\varepsilon}\big) = P\big(\vect{\varepsilon},\varphi \big)$, normalized,
$\sum_{\vect{\varepsilon}} \int_a^b \ud \varphi P\big(\varphi,\vect{\varepsilon}\big) = 1$,
and can be written as
\be \label{Eq:conditional2}
P\big(\vect{\varepsilon},\varphi \big) = P\big(\vect{\varepsilon} \vert \varphi \big) P\big(\varphi \big),
\ee 
where $P\big(\vect{\varepsilon} \vert \varphi \big)$ is the conditional probability of observing 
$\vect{\varepsilon}$ given $\varphi$, and $P\big(\varphi \big)$ is the prior probability density function.
Analogously we can write 
\be \label{Eq:conditional3}
P\big(\varphi,\vect{\varepsilon}\big) = P\big(\varphi \vert \vect{\varepsilon} \big) P\big(\vect{\varepsilon}\big),
\ee 
where $P\big(\varphi \vert \vect{\varepsilon}\big)$ is the posterior probability density function
given the measurement results $\vect{\varepsilon}$ and $P(\vect{\varepsilon})$ is known as marginal likelihood.
By combining Eqs.~(\ref{Eq:conditional2}) and (\ref{Eq:conditional3}), 
we obtain the Bayes' theorem \cite{Bayes}:
\be \label{Bayes_invert}
P\big(\varphi \vert \vect{\varepsilon}\big) = \frac{P\big(\vect{\varepsilon} \vert \varphi\big) P\big(\varphi \big)}{P\big(\vect{\varepsilon}\big)}, 
\ee
where $P\big(\vect{\varepsilon}\big) = \int_a^b \ud \varphi P\big(\vect{\varepsilon} \vert \varphi \big) P\big(\varphi \big)$ 
provides the normalization of the posterior probability density function.
As already mentioned, $\varphi$ is a random phase variable that changes in different sets of $m$ data
with probability $P\big(\varphi \big)$. 
In a typical phase estimation problem, however, the phase is not a random variable -- it has a fixed unknown value -- and in this context 
the Bayes' theorem is of little use. 
Here enters the Bayesian probability interpretation. 
Now the prior is not considered as the probability distribution of a random variable, but as a measure of our
ignorance about the (fixed) true value of the phase shift. Once we subjectively fix the prior 
probability $P(\varphi = \theta)$ (the probability that $\varphi$ is equal to the true value of the phase shift $\theta$) 
on the base of the available knowledge before making any measurement (for instance, we might know that $\theta$ is different from zero only in a small interval) 
we can obtain from the Bayes' theorem the posterior probability distribution $P\big(\varphi \vert \vect{\varepsilon}\big)$. 
In other words, the objective knowledge gained by the measurements updates 
the initial prior which was chosen on a subjective base choice.
This is working tool of the Bayesian inference, whose main features are 
discussed below.


\subsubsection{Point estimates}

-- The posterior probability density function contains all the available statistical (Bayesian) information about $\theta$
given the measurement results $\vect{\varepsilon}$. 
It is often useful to obtain from it a value, the estimator or point estimate, which represents 
the best guess about the true value of the phase shift.
The most popular point estimates are the posterior mean,
\be
\bar{\Theta}(\vect{\varepsilon}) = \int_a^b \ud \varphi \, \varphi \, P(\varphi \vert \vect{\varepsilon}), 
\ee
and the absolute maximum, 
\be
\Theta_{\rm max}(\vect{\varepsilon}) = {\rm arg} \Big[ \max_{\varphi \in [a,b]} P(\varphi \vert \vect{\varepsilon}) \Big].
\ee
With a flat prior, $\Theta_{\rm max}(\vect{\varepsilon}) $ coincides with the maximum likelihood estimator (see Sec.~\ref{MaxLik}).


\subsubsection{Confidence intervals}

-- From $P(\varphi \vert \vect{\varepsilon})$ we can compute the probability that 
the parameter $\varphi$ lies in a particular 
region $\Omega$ of the parameter space:
\be \label{ConfInt}
P(\varphi \in \Omega \vert \vect{\varepsilon}) = \int_\Omega \ud \varphi \, P(\varphi \vert \vect{\varepsilon}).
\ee
For instance, we might be interested to find the smallest region $\Omega$ for a given value 
of $P(\varphi \in \Omega \vert \vect{\varepsilon})$
or, as often done in practice, calculate the confidence (or credible) interval $2 \Delta$, 
\be
P(\varphi \in 2\Delta \vert \vect{\varepsilon}) = \int_{\Theta_{\vect{\varepsilon}} - \Delta}^{\Theta_{\vect{\varepsilon}} + \Delta} \ud \varphi \, P(\varphi \vert \vect{\varepsilon}).
\ee
around a certain estimate $\Theta_{\vect{\varepsilon}}$ of the phase shift.
In the case of a Gaussian posterior distribution centred in $\Theta_{\vect{\varepsilon}}$ and of variance $\sigma^2$, 
the $68.27\%$ confidence interval corresponds to a standard deviation $\Delta=\sigma$, 
the $95.45\%$ confidence interval corresponds to two standard deviations $\Delta=2\sigma$,
etc. 
Finally, it is often useful to calculate 
\be \label{posteriorvariance}
(\Delta \varphi)^2_{\vect{\varepsilon}} \equiv \int_{a}^b \ud \varphi \, P(\varphi \vert \vect{\varepsilon}) \, \big(\varphi - {\Theta}_{\vect{\varepsilon}}\big)^2,
\ee
called the posterior variance.


\subsection{Large sample properties}
\label{BayesChapter:largesample}

Asymptotically in the sample size, the posterior becomes normally distributed and centred at the true value of the parameter. 
This is a consequence of the Laplace-Bernstein-von Mises theorem \cite{LehmannBOOK1998, LeCamBOOK1986} 
and will be discussed in the following.

Let us assume that $P(\varphi \vert \vect{\varepsilon})$ is nonzero and has continuous derivatives.
Since the logarithm is monotone, the maximum of $P(\varphi \vert \vect{\varepsilon} )$ is also
the maximum of $\ln P(\varphi \vert \vect{\varepsilon} )$. 
We thus expand $\ln P(\varphi \vert \vect{\varepsilon} )$ in 
Taylor series around $\Theta_{\rm max}$ and then take the exponential. We have
\be \label{BayNormEq2}
P\big(\varphi \vert \vect{\varepsilon}\big) = \exp\bigg\{ \ln P(\Theta_{\rm max} \vert \vect{\varepsilon} )   
+ \sum_{k \geq 2} \frac{1}{k!} \frac{d^k \ln P(\varphi \vert \vect{\varepsilon} )}{d \varphi^k}\Big\vert_{\Theta_{\rm max}}   (\varphi - \Theta_{\rm max})^k \bigg\},
\ee
where we have taken into account that $d \ln P(\varphi \vert \vect{\varepsilon} )/d \varphi\vert_{\Theta_{\rm max}}=0$.
The zeroth order term of the expansion, $\exp \{ \ln P(\Theta_{\rm max} \vert \vect{\varepsilon} ) \}$, 
can be absorbed in the normalization of $P\big(\varphi \vert \vect{\varepsilon}\big)$.
We thus have have 
\beq \label{BayNormEq3}
P\big(\varphi \vert \vect{\varepsilon}\big)  \propto 
\exp\bigg\{ - \frac{\mathcal{I}(\vect{\varepsilon})}{2} (\varphi - \Theta_{\rm max})^2 \bigg\} 
\prod_{k> 2} 
\exp\bigg\{ \frac{1}{k!} \frac{d^k \ln P(\varphi \vert \vect{\varepsilon} )}{d \varphi^k}\Big\vert_{\varphi=\Theta_{\rm max}}   (\varphi - \Theta_{\rm max})^k \bigg\} \nonumber \\
\eeq
where 
\be \label{BayNormEq4}
\mathcal{I}(\vect{\varepsilon}) \equiv - \frac{d^2 \ln P(\varphi \vert \vect{\varepsilon} )}{d \varphi^2}\Big\vert_{\Theta_{\rm max}}.
\ee
Since $\Theta_{\rm max}$ is the maximum of $P(\varphi \vert \vect{\varepsilon} )$ we have that $\mathcal{I}(\vect{\varepsilon})$ 
is strictly positive and, in particular, non-vanishing.
Let us now consider $m$ independent measurements with results $\vect{\varepsilon}=\{\varepsilon_1, \varepsilon_2, \ldots, \varepsilon_m\}$.
In this case the posterior probability density function is, up to a normalization constant, 
$P\big(\varphi \vert \vect{\varepsilon}\big) \propto \prod_{i=1}^m P(\varphi \vert \varepsilon_i)$.
We can also rewrite it as 
$P\big(\varphi \vert \vect{\varepsilon}\big) \propto \prod_{\varepsilon} \big[P(\varphi \vert \varepsilon )\big]^{m \times \frac{m_{\varepsilon}}{m}}$, 
where the product runs over all possible values of $\varepsilon$ and $m_{\varepsilon}/m$
is the observation frequency of the result $\varepsilon$ in $m$ measurements.
Asymptotically in $m$, the frequency tends to the probability
$m_{\varepsilon}/m \to P(\varepsilon \vert \theta)$, where $\theta$ is the true value of the phase. We get
\be \label{BayNormEq5}
\frac{\ln P\big(\varphi \vert \vect{\varepsilon}\big)}{m} \to \sum_{\varepsilon} P(\varepsilon \vert \theta) \ln P\big(\varphi \vert \varepsilon \big) + {\rm const},
\ee
where the constant term is due to the normalisation of $P\big(\varphi \vert \vect{\varepsilon}\big)$.
We now compute Eq.~(\ref{BayNormEq4}) using Eq.~(\ref{BayNormEq5}).
We obtain $\mathcal{I}(\vect{\varepsilon}) = m \mathcal{F}$, where
\footnote{
Note that $\mathcal{F}$ does not depend on $\vect{\varepsilon}$.
Where did the dependence go ? The fact is that, in the limit $m\to \infty$, all the possible sequences 
$\vect{\varepsilon}=\varepsilon_1, ..., \varepsilon_m$ converge to a ``typical sequence'', differing only in the order of $\varepsilon_i$,   
such that the outcome frequency $m_\varepsilon/m$ of each possible measurement result $\epsilon$ converges to the corresponding probability $P(\varepsilon \vert \theta)$.
} 
\be
\mathcal{F} =  - \sum_{\varepsilon} P(\varepsilon \vert \theta) \frac{\partial^2 \ln P\big(\varphi \vert \varepsilon \big)}{\partial \varphi^2}\bigg\vert_{\Theta_{\rm max}}.
\ee
We have thus found that the Gaussian term in Eq.~(\ref{BayNormEq3}) has a width proportional to $1/\sqrt{m}$.
In the phase interval $\vert \varphi - \Theta_{\rm max} \vert \sim 1/\sqrt{m}$
higher order $k>2$ terms in Eq.~(\ref{BayNormEq3}) give a contribution 
$\propto 1/m^{k/2-1}$ and are thus negligible when $m \to \infty$.
To the leading order in $m$, we thus have 
\be
P\big(\varphi \vert \vect{\varepsilon}\big)  \propto 
\exp\bigg( - \frac{m \mathcal{F}}{2} (\varphi - \Theta_{\rm max})^2 \bigg),  \qquad {\rm for\,\,}m \gg 1.
\ee
In the following we clarify the relation between $\Theta_{\rm max}$ and $\mathcal{F}$ with 
the true value of the phase shift, $\theta$ and the FI, respectively. 
By taking the derivative of Eq.~(\ref{BayNormEq5}) with respect to $\varphi$, we have 
\beq \label{Bayes:expansion:1}
\frac{d \ln P\big(\varphi \vert \vect{\varepsilon}\big)}{d \varphi} =
m \sum_\varepsilon \frac{P(\varepsilon \vert \theta)}{P(\varphi \vert \varepsilon)} \frac{d P(\varphi \vert \varepsilon)}{d \varphi}      
= m \bigg( \sum_\varepsilon \frac{P(\varepsilon \vert \theta)}{P(\varepsilon \vert \varphi) } \frac{d P(\varepsilon \vert \varphi)}{d \varphi} +\frac{d P(\varphi)}{d \varphi}  \bigg), \nonumber \\
\eeq
where we used the Bayes' theorem, $P(\varphi \vert \varepsilon) = P(\varepsilon \vert \varphi) P(\varphi)/P(\varepsilon)$. 
If $\frac{d P(\varphi)}{d \varphi}\vert_{\theta}=0$, Eq.~(\ref{Bayes:expansion:1}) shows that the maximum of the 
posterior probability density function has, asymptotically in $m$, a maximum at the true value of the parameter.
In the special case of a flat prior we have that 
$\Theta_{\rm max}=\Theta_{\rm ML}$ and the above result is equivalent to the consistency of the 
maximum likelihood estimator, discussed in Sec.~\ref{ConsML}.
Taking the second derivative of Eq.~(\ref{BayNormEq5}) we find
\be
\mathcal{F} = \sum_\varepsilon  \frac{1}{P(\varepsilon \vert \varphi)} \bigg(\frac{\ud P(\varepsilon \vert \varphi)}{\ud \varphi}\bigg)^2 \bigg\vert_{\theta} 
-  \frac{d^2P(\varphi)}{d\varphi^2}\bigg\vert_{\theta}.
\ee
This shows that, if $d^2P(\varphi)/d\varphi^2\vert_{\theta}=0$, then $\mathcal{F}$ exactly
coincides with the FI, Eq.~(\ref{FisherInformation}), calculated at the true value of the parameter: $\mathcal{F} = F(\theta)$.
To the leading order in $m$, 
we thus have 
\be \label{Bayes:expansion:2}
P\big(\varphi \vert \vect{\varepsilon}\big) = \sqrt{\frac{m F(\theta)}{2 \pi}} \, e^{ - \frac{m F(\theta)}{2} (\theta - \varphi)^2},
\qquad {\rm for\,\,}m \gg 1.
\ee 
We recall that this results requires $d P(\varphi)/d\varphi \vert_{\theta}=0$ and $d^2P(\varphi)/d\varphi^2\vert_{\theta}=0$,
which are clearly satisfied for a flat prior. 


\subsection{Bounds for posterior variance}
\label{BayesChapter:bounds}

Here we derive a bound for the posterior variance Eq.~(\ref{posteriorvariance}),
calculated around the arbitrary point ${\Theta}_{\vect{\varepsilon}}$ 
and valid for any $\vect{\varepsilon}$. 
Assuming that $P(\varphi \vert \vect{\varepsilon})$ vanishes at the borders of the phase domain, 
we have \footnote{
We here demonstrate a generalised form of Eq.~(\ref{Eq.BayCR})  \cite{GhoshSPL1993}.
Let $ \gamma(\varphi)$ be an arbitrary parametric differentiable function of $\varphi$
and $\gamma'(\varphi) = \frac{\partial \gamma(\varphi)}{\partial \varphi}$ its derivative.
Under the conditions $\lim_{\varphi \to a} P(\varphi \vert \vect{\varepsilon}) = \lim_{\varphi \to b} P(\varphi \vert \vect{\varepsilon})$
and $\lim_{\varphi \to a} \gamma(\varphi) P(\varphi \vert \vect{\varepsilon}) = \lim_{\varphi \to b} \gamma(\varphi)P(\varphi \vert \vect{\varepsilon})$, 
 we have 
\beq \label{Proof.2} 
\int_a^b \ud \varphi \frac{\partial P(\varphi \vert \vect{\varepsilon})}{\partial \varphi } = 0. \nonumber
\eeq
and 
\beq \label{Proof.1}
\int_a^b \ud \varphi \, \gamma(\varphi) \, \frac{\ud P(\varphi \vert \vect{\varepsilon})}{\ud\varphi} = - \int_a^b \ud \varphi \, \gamma'(\varphi) P(\varphi \vert \vect{\varepsilon}), \nonumber
\eeq
respectively.
Note that the above limit conditions on $P(\varphi \vert \vect{\varepsilon})$ are both fulfilled if $\lim_{\varphi \to a,b} P(\varphi \vert \vect{\varepsilon}) = 0$
which is obtained, for instance, if the prior vanishes at the border of the phase domain, $\lim_{\varphi \to a,b} P(\varphi)=0$.
We thus obtain 
\beq \label{Proof.3} 
\int_a^b \ud \varphi \, \big(\gamma(\varphi) - \lambda_{\vect{\varepsilon}} \big)  \, \frac{\ud P(\varphi \vert \vect{\varepsilon})}{\ud\varphi} = 
- \int_a^b \ud \varphi \, \gamma'(\varphi) P(\varphi \vert \vect{\varepsilon}), \nonumber
\eeq
where $\lambda_{\vect{\varepsilon}}$ is a $\varphi$-independent generic real function of $\vect{\varepsilon}$.
By taking the square of the above equation, we obtain 
\beq
\bigg( \int_a^b \ud \varphi \, \gamma'(\varphi) P(\varphi \vert \vect{\varepsilon}) \bigg)^2 
\leq 
\int_a^b \ud \varphi \, \frac{1}{P(\varphi \vert \vect{\varepsilon})}
\bigg( \frac{\ud P(\varphi \vert \vect{\varepsilon})}{\ud\varphi} \bigg)^2
\int_a^b \ud \varphi \, P(\varphi \vert \vect{\varepsilon})\, \big(\gamma(\varphi) - \lambda_{\vect{\varepsilon}} \big)^2, \nonumber
\eeq
as follows from a Cauchy-Schwarz inequality.
Equation~(\ref{Eq.BayCR}) is recovered for $ \lambda_{\vect{\varepsilon}} =  \Theta_{\vect{\varepsilon}}$ and $\gamma(\varphi)=\varphi$.
In this case, the equality sign in Eq.~(\ref{Eq.BayCR}) is obtained if and only if 
$\frac{\ud}{\ud\varphi} \log P(\varphi \vert \vect{\varepsilon}) = c_{\vect{\varepsilon}} \, (\varphi - {\Theta}_{\vect{\varepsilon}})$,
where $c_{\vect{\varepsilon}}$ is a positive function of $\vect{\varepsilon}$, i.e. if and only if $P(\varphi \vert \vect{\varepsilon})$ is a 
Gaussian centered in ${\Theta}_{\vect{\varepsilon}}$ and 
variance $1/c(\vect{\varepsilon})$.
 }
\be \label{Eq.BayCR}
(\Delta \varphi)_{\vect{\varepsilon}}^2 \geq \frac{1}{G(\vect{\varepsilon})},
\ee
where 
\be \label{Eq.G}
G(\vect{\varepsilon}) \equiv
\int_a^b \ud \varphi \frac{1}{P(\varphi \vert \vect{\varepsilon})} \bigg( \frac{\ud P(\varphi \vert \vect{\varepsilon})}{\ud \theta} \bigg)^2. 
\ee
The bound~(\ref{Eq.BayCR}) presents interesting differences with respect to the Cram\'er-Rao bound (see Sec.~\ref{crlb}).
It depends on the measured results $\vect{\varepsilon}$ and does not explicitly depend on the bias of the estimator. 
Differently from the FI, Eq.~(\ref{Eq.G}) is not additive, as a consequence of the 
normalisation of $P(\varphi \vert \vect{\varepsilon})$.
In addition, the above theorem, can be generalized to arbitrary parametric functions of $\varphi$ [see footnote $(^{19})$].
Asymptotically in the number of measurements and for a flat prior, Eq.~(\ref{Bayes:expansion:2})
holds, and thus $G(\vect{\varepsilon})= mF(\theta)$. 
Equation~(\ref{Eq.BayCR}) thus reduce to:
\be
(\Delta \varphi)_{\vect{\varepsilon}}^2 \geq 
\frac{1}{m \, F(\theta)}, \qquad {\rm for \,\, } m\gg1.
\ee
We can further calculate the statistical average of Eq.~(\ref{Eq.BayCR}), 
$(\Delta \varphi)^2 \equiv \sum_{\vect{\varepsilon}} (\Delta \varphi)_{\vect{\varepsilon}}^2 P(\vect{\varepsilon} \vert \theta)$.
By using Jensen's inequality~\footnote{
We use the facts that {\it i}) the $G(\vect{\varepsilon})$ is strictly positive, 
{\it ii}) $\sum_{\vect{\varepsilon}} P(\vect{\varepsilon} \vert \theta) = 1$ 
and {\it iii}) the function $1/x$ is convex for $x >0$.
}, we find
\be \label{Eq.G2}
(\Delta \varphi)^2 \geq \frac{1}{\sum_{\vect{\varepsilon}} P(\vect{\varepsilon} \vert \theta) G(\vect{\varepsilon})}.
\ee
In general the bound~(\ref{Eq.G2}) is not analytical but can be evaluated with a MonteCarlo method.

\section{Quantum interferometry and entanglement}
\label{QI&ENT}

Entanglement is an experimentally-demonstrated fundamental property of Nature 
(for reviews see \cite{HorodekiRMP2009, GuhnePHYSREP2009}).
However, for many decades after the birth of quantum mechanics, entanglement was considered a nuisance, 
admittedly an esoteric one \cite{ZurekBOOK}. 
Only in the last thirty years it has been recognised that entanglement can be an asset to 
tremendously improve the performances of certain classical tasks \cite{NielsenBOOK}. 
Entanglement-based technologies span from secure communication to 
the speed-up of factorization algorithms passing through the possibility
to ``play tricks that cannot be imitated by classical magicians'' \cite{BrussJMP2002}.
However, we still miss the deep physical reason of why entanglement is a resource. 
We do not know which specific characteristics of quantum correlations are really crucial to overcome specific classical protocols. 

In the context of phase estimation, the idea that quantum correlations are {\it necessary} to overcome the 
shot noise sensitivity $\Delta \theta_{\rm SN} = 1/\sqrt{N m}$, where $N$ is the number of qubits and 
$m$ is the number of measurements, emerged already in early pioneer works \cite{Yurke_1986, WinelandPRA1992,WinelandPRA1994}.
Major steps forward were the recognition that the shot noise can be overcome by an experimentally valuable 
class of states, known as spin squeezed states \cite{WinelandPRA1992, WinelandPRA1994}, 
and that such states are entangled~\cite{SorensenNATURE2001}.
However, spin squeezed states are a relatively small class of useful states. 
The prominent counter-example are maximally entangled states (``Schr\"odinger-cat'', NOON or GHZ states, see below), which are not spin squeezed and 
nevertheless can provide a sub shot noise phase sensitivity~\cite{BollingerPRA1996, HuelgaPRL1997}.  
It was finally demonstrated in Ref.~\cite{GiovannettiPRL2006}, in a quantum information framework, 
that separable states can be exploited to reach, at best, the shot noise.    
It was also shown that the maximum allowed phase sensitivity (achievable by maximally entangled states) is the 
Heisenberg limit $\Delta \theta_{\rm HL} = 1/N\sqrt{m}$.
Yet, not all entangled states are equally useful:
only a special class of quantum correlations can be exploited in interferometry to
estimate a phase shift with sensitivity higher than the classical shot-noise.
This special class of useful entangled states is fully recognised \cite{PezzePRL2009} and quantified 
\cite{HyllusPRA2012,TothPRA2012} by the Fisher information, $F$.
The condition $F \geq N$ is sufficient for entanglement and {\it necessary and sufficient} for a state to 
be useful to achieve a sub shot noise sensitivity \cite{PezzePRL2009}.

We start this section by giving the (mathematical) definition of separability/entangled. We later demonstrate that the Fisher information provides
an entanglement criterion that depends on the quantum state but also on the Hamiltonian generating the phase shift transformation
and the choice of the observable. 
It is quite obvious that having in our hand a criterion to recognize the useful entanglement for a phase estimation protocol is crucially important when trying to build an interferometer.  But even more importantly, it provides the physical reason why entangled states can be useful and which physical characteristic they should contain for reaching sub-shot noise sensitivities. 

\subsection{Composite systems and entanglement}

Let us consider two  {\it independent} (non-interacting) physical systems, 
say A and B, with Hilbert state space $\mathcal{H}_{\rm A}$ and $\mathcal{H}_{\rm B}$, respectively. 
The composite state space is the tensor product $\mathcal{H}_{\rm AB} = \mathcal{H}_{\rm A} \otimes \mathcal{H}_{\rm B}$.
If system A is in the pure state $\ket{\psi_{\rm A}}$ and system B in the pure state $\ket{\psi_{\rm B}}$, the composite
system is in state 
\be \label{pure_separable}
\ket{\psi_{\rm sep}} = \ket{\psi_{\rm A}} \otimes \ket{\psi_{\rm B}}.
\ee
States which can be represented in this form are called 
product (or separable) states:  any {\it local operation} acting on A would not affect B and viceversa. 
As a consequence, the expectation value of any joint measurement
$\hat{M}_{\rm AB} = \hat{M}_{\rm A} \otimes \hat{M}_{\rm B}$ done in $\mathcal{H}_{\rm AB}$
is equal to the product of expectation values calculated in each subsystem: 
$\bra{ \psi_{\rm AB} } \hat{M}_{\rm AB} \ket{ \psi_{\rm AB} } =  
\bra{ \psi_{\rm A} } \hat{M}_{\rm A} \ket{ \psi_{\rm A} } \times 
\bra{ \psi_{\rm B} } \hat{M}_{\rm B} \ket{ \psi_{\rm B} }.$
Even if the two systems A and B are independent, they can still communicate with each others. 
This communication can create classical correlations, with A and B being
now in a state $\ket{\psi_{\rm A}^{(k)}} \otimes \ket{\psi_{\rm B}^{(k)}}$ with some 
probability $p_k$ [with $p_k>0$ and $\sum_k p_k=1$].
Therefore, a (mixed) state of a composite quantum system is called classically correlated (or separable)
if it can be written as a convex combinations of separable pure state density matrices \cite{WernerPRA1989, PeresPRL1996},
\be
\hat{\rho}_{\rm sep} = \sum_k p_k \ket{\psi^{(k)}_{\rm A}} \bra{\psi^{(k)}_{\rm A}} \otimes \ket{\psi^{(k)}_{\rm B}} \bra{\psi^{(k)}_{\rm B}}.
\ee
States that are not classically correlated are called entangled.
Local operations and classical communication (LOCC) cannot create or destroy entanglement.  
The definition of separability/entanglement can be straightforwardly extended to the
case $N$ composite subsystems. A state in
$ \mathcal{H}_1 \otimes \mathcal{H}_2 \otimes ... \otimes \mathcal{H}_N$
is said to be separable if it can be written as
\be \label{seps}
\hat{\rho}_{\rm sep} = \sum_k p_k 
\ket{\psi_1^{(k)}}\bra{\psi_1^{(k)}} \otimes 
\ket{\psi_2^{(k)}}\bra{\psi_2^{(k)}} \otimes ... \otimes
\ket{\psi_N^{(k)}}\bra{\psi_N^{(k)}}.
\ee
States that cannot be written as Eq.~(\ref{seps}) are entangled.

\subsection{Entanglement and phase sensitivity}

In this section we discuss the relation between entanglement and 
phase sensitivity. We consider:
\begin{itemize}

\item a finite number, $N$, of distinguishable subsystems (e.g. distinguishable particles) labeled by $i=1,2,...,N$.
The Hilbert space of each subsystem, $\mathcal{H}_i$, has an arbitrary, finite, dimension.
The case of $N$ qubits (${\rm dim}\mathcal{H}_i=2$ $\forall i$) is a relevant example 
which will be discussed in more details in Sec.~\ref{SU2};

\item a local and unitary phase shift operation, $e^{-i \theta \hat{h}_i}$, on each subsystem. 
Here $\hat{h}^{(i)}$ is the (local) generator of the phase shift;

\item the same phase shift $\theta$ equal for all the $N$ subsystems.
The general transformation acting on the $N$ particles is thus $\otimes_{i=1}^N e^{-i \hat h_i \theta} = e^{-i \hat H \theta} $,  where
$\hat{H} = \sum_{i=1}^N \hat{h}_i$.

\end{itemize}
In the following we assume, for simplicity, that 
all $N$ subsystems have the same Hilbert space dimension and $\hat{h}^{(i)}$ is the same ($\hat{h}^{(i)} = \hat h$) for all of them.
We indicate with $h_{\rm max}$ and $h_{\rm min}$ the maximum and minimum eigenvalues of $\hat{h}$, respectively, 
and with $\vert h_{\rm max} \rangle$ and $\vert h_{\rm min} \rangle$ the corresponding eigenvectors.
We will discuss separately the three cases (for a pictorial representation see Fig.~\ref{Fig:GLM}): 
separable [Fig.~\ref{Fig:GLM}(a), Sec.~\ref{Sec:separable}], 
general multiparticle entangled [Fig.~\ref{Fig:GLM}(b), Sec.~\ref{Sec:kentangled}] and 
maximally entangled [Fig.~\ref{Fig:GLM}(c), Sec.~\ref{Sec:entangled}] states.

\begin{figure}[h!]
\begin{center}
\includegraphics[clip,scale=2]{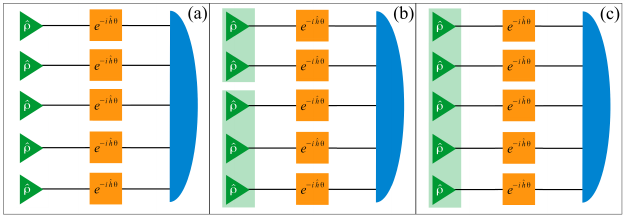}
\end{center}
\caption{{\bf Different scenarios in quantum metrology.} 
The $N$ particles of the probe (green triangles) are prepared in a separable (a),
multiparticle entangled (b) and maximally entangled (c) state.
The orange boxes are unitary transformations on each particle.
The shaded green region indicates entanglement between the input particles.
The blue D-shape indicate general (separable or entangled) detection.
} \label{Fig:GLM}
\end{figure}

\subsubsection{Separable states}
\label{Sec:separable}

-- Here we show that for 
 $\hat{\rho}_{\rm sep} = \sum_k p_k \ket{\psi_{\rm sep}^{(k)}}\bra{\psi_{\rm sep}^{(k)}}$ with $\ket{\psi_{\rm sep}^{(k)}} = \ket{\psi_1^{(k)}} \otimes \ket{\psi_2^{(k)}}  \otimes ... \otimes \ket{\psi_N^{(k)}}$,
see Eq.~(\ref{seps}), the FI is strictly bounded~\cite{GiovannettiPRL2006}.
We have the following chain of inequalities: 
 \beq \label{Fishsepsep}
F_Q\big[\hat{\rho}_{\rm sep}\big]
\leq
\sum_k p_k  F_Q\big[\ket{ \psi_{\rm sep}^{(k)} } \big]
= \sum_{i, k} p_k F_Q\big[\ket{ \psi_{i}^{(k)} } \big]
\leq N \big( h_{\rm max} - h_{\rm min} \big)^2.  \nonumber \\
\eeq
The first inequality comes from the convexity of the QFI (see Sec.~\ref{convexityf}),
the middle equality is a consequence of the additivity of the QFI (see Sec.~\ref{additivityf}) and 
the last inequality follows from $F_Q\big[\ket{ \psi_{i}^{(k)} } \big] \leq 4 (\Delta \hat h_i)^2$ 
(see Sec.~\ref{unitaryQFI} )
and $4 (\Delta \hat h_i)^2 \leq (h_{\rm max} - h_{\rm min})^2$.
This brings us to an important result: the optimal phase sensitivity 
(the quantum Cramer-Rao bound) for separable states of $N$ particle is
\be \label{SNbound}
\Delta \theta_{\rm SN} =  \frac{1}{\sqrt{ N m} } \frac{1}{\vert h_{\rm max} - h_{\rm min}\vert  },
\ee 
independently of the specific measurement and estimator. 
The bound~(\ref{SNbound}) is generally indicated as the shot noise (or standard quantum) limit.
In this bound, the number of particles plays the role of a statistical gain, in complete analogy 
to the number of independent repetitions of the measurements. 
Also, going from qubits (in which case $h_{\rm max} - h_{\rm min}=1$) to more complex (multimode) systems, 
the shot noise decreases by a factor proportional to the number of modes. 

To illustrate this result, let us consider the probe state 
\be
\ket{\psi_{\rm sep}} = \bigg( \frac{\ket{h_{\rm max}} + \ket{h_{\rm min}} }{\sqrt{2}} \bigg)^{\otimes N}.
\ee
Since the the phase shift operator is 
$e^{-i \hat{H} \theta} = \otimes_{i=1}^N e^{-i \hat{h}^{(i)} \theta}$,
the output state becomes
\be \label{sepPSstate}
\ket{\psi_{\rm sep}(\theta)} = \bigg( \frac{ e^{-i h_{\rm max}\theta} \ket{h_{\rm max}} + e^{-i h_{\rm min}\theta} \ket{h_{\rm min}} }{\sqrt{2}} \bigg)^{\otimes N}.
\ee
To estimate the phase shift, we can take a separable POVM with two elements: the projection over the probe state, which has probability 
\be \label{sepProb}
\big| \langle \psi_{\rm sep} | \psi_{\rm sep}(\theta) \rangle \big|^2 = \cos^{2 N} \bigg( \frac{\theta ( h_{\rm max} - h_{\rm min} )}{2} \bigg),
\ee
and the projection over the subspace orthogonal to the probe state, which has probability $1 - | \langle \psi_{\rm sep} | \psi_{\rm sep}(\theta) \rangle|^2$. 
The FI can be calculated straightforwardly (see Sec.~\ref{QFIgeneral}) and, in the limit $\theta \to 0$, it saturates the right hand side of Eq.~(\ref{Fishsepsep}).
The CRLB is thus $\Delta \theta_{\rm CR}  = 1/\sqrt{N m} \vert h_{\rm max} - h_{\rm min} \vert$, Eq.~(\ref{SNbound}).
To have an insight on this result, let us notice that, for $N\gg 1$, 
$| \langle \psi_{\rm sep} | \psi_{\rm sep}(\theta) \rangle|^2 \approx e^{-N \big(\frac{ h_{\rm max} - h_{\rm min}}{2} \big)^2 \theta^2}$.
The input state and the phase-shifted one thus become ``distinguishable'' 
for a phase shift $\theta \propto 1/\sqrt{N} \vert h_{\rm max} - h_{\rm min} \vert$.
This intuitively is the smallest phase shift detectable and coincides with the QCR.

\subsubsection{Entangled probe}
\label{Sec:entangled}

-- From the above results, the condition 
\be \label{Fishent}
F\big[\hat{\rho}\big] > N \big( h_{\rm max} - h_{\rm min}\big)^2,
\ee
is sufficient for entanglement, i.e. it cannot be fulfilled if $\hat \rho$ is a separable state.
To be more precise, the inequality~(\ref{Fishent}) is the condition 
for {\it useful entanglement}: it is necessary and sufficient for a state to be useful to estimate a phase shift $\theta$
with a sensitivity overcoming the bound for separable states, 
Eq.~(\ref{SNbound}).
The crucial point to notice is that not all entangled states are useful \cite{PezzePRL2009}:
the useful ones are singled out by their FI.

Here we investigate the maximum sensitivity achievable with 
an entangled probe state \cite{GiovannettiPRL2006}.
By using the convexity of the QFI
($F_Q\big[\sum_k \gamma_k \vert \psi_k \rangle \langle \psi_k \vert\big] \leq \sum_k \gamma_k F_Q\big[\vert \psi_k \rangle\big] 
\leq \max_{\ket{\psi}} F_Q\big[ \ket{\psi}\big]$, 
where the maximum is taken over all pure states and $\sum_k \gamma_k=1$),
we have 
\be \label{Fishentent}
F_Q\big[\hat{\rho}\big] 
\leq \max_{\ket{\psi}} F_Q\big[ \ket{\psi} \big] \leq N^2 \big( h_{\rm max} - h_{\rm min}\big)^2.
\ee 
The last inequality is obtained by noticing that 
$\max_{\ket{\psi}} F_Q\big[ \ket{\psi} \big] = 4 \max_{\ket{\psi}} \big( \Delta \hat{H} \big)^2_{\ket{\psi}}$, 
where the maximum variance is given by the difference between the largest, $H_{\rm max}$,  
and the smallest, $H_{\rm min}$, eigenvalue of the collective Hamiltonian $\hat{H}$, 
$( \Delta \hat{H})^2 \leq (H_{\rm max} - H_{\rm min})^2/4$.
Since the Hamiltonian $\hat{H}$ is linear, we further have 
$H_{\rm max} = N h_{\rm max} $ and $H_{\rm min} = N h_{\rm min}$.
Taking all these inequalities together, we find
$F\big[\hat{\rho}\big] \leq N^2 \big( h_{\rm max} - h_{\rm min}\big)^2$ and thus Eq.~(\ref{Fishentent}).
We arrive at a second important result: the maximum phase sensitivity allowed by quantum mechanics, 
the Heisenberg limit (HL), is given by 
\be \label{HLbound}
\Delta \theta_{\rm HL} =  \frac{1}{N \sqrt{m}} \frac{1}{\vert h_{\rm max} - h_{\rm min} \vert }.
\ee
The difference between Eq.~(\ref{SNbound}) and Eq.~(\ref{HLbound}) is a faster scaling 
of phase sensitivity with the number of particles, which cannot be obtained by classical means
\footnote{
To be more precise, a scaling of phase sensitivity $\Delta \theta \propto 1/N$ can be indeed obtained by classical means
if one trades physical resources (entanglement) with running time, for instance by applying the same phase shift $N$ times 
to a single qubit \cite{RudolphPRL2003, DeBurghPRA2005, VanDamPRL2007}: a technique known as multi-pass interferometry \cite{HiggingNATURE2007, BerryPRA2009},
see also \cite{ReschPRL2007, AfekPRL2010}.
}.
The HL can be saturated by 
\be \label{MaxEnt}
\ket{\psi_{\rm GHZ}} = \frac{
\ket{h_{\rm max}}^{\otimes N}  + 
\ket{h_{\rm min}}^{\otimes N} }{\sqrt{2}}, 
\ee
which is a maximally entangled state in the basis of eigenstates of $\hat{h}$.
The state Eq.~(\ref{MaxEnt}) is often referred to as ($N$-substems) Greenberger-Horne-Zeilinger (GHZ) \cite{GHZ}
or ``Schr\"odinger-cat'' state.
To see that $\ket{\psi_{\rm GHZ}}$ saturates the HL, let us apply the unitary transformation $e^{-i \hat{H} \theta} = \otimes_{i=1}^N e^{-i \hat{h}_i \theta}$, to obtain 
\be
\ket{\psi_{\rm GHZ}(\theta)} = \frac{
e^{-i h_{\rm max} N \theta} \ket{h_{\rm max}}^{\otimes N}  + 
e^{-i h_{\rm min} N \theta} \ket{h_{\rm min}}^{\otimes N} }{\sqrt{2}}. 
\ee
A comparison with Eq.~(\ref{sepPSstate}) reveals that, by applying the phase shift operator to the maximally entangled state, 
the phase shift is ``amplified'' by the total number of particles.
Taking an output measurement with entangled POVM of elements given by 
the projection over the state $\ket{\psi_{\rm GHZ}(\theta)}$ and the orthogonal subspace,
we find 
\be
\vert \langle \psi \vert \psi(\theta) \rangle \vert^2 = \cos^2 \bigg( \frac{N \theta (h_{\rm max}-h_{\rm min})}{2} \bigg),
\ee
and $1- \vert \langle \psi \vert \psi(\theta) \rangle \vert^2 = \sin^2(N \theta (h_{\rm max}-h_{\rm min})/2)$.
These probabilities oscillates in phase $N$ time faster than corresponding 
probabilities for the separable state, see Eq.~(\ref{sepProb}).
Distinguishability between the probe state and the phase-shifted one
is first reached when the phase shift is $\theta \propto 1/N \vert h_{\rm max} - h_{\rm min} \vert$.
This coincides with the smallest possible detectable phase shift and Cramer-Rao bound, 
as shown by an explicit calculation of the FI \footnote{
The same result can be obtained by the separable POVM of elements
\beq
\hat{\Pi}_0 = \otimes_{i=1}^N \Big( \ket{h^{(i)}_{\rm max}}\bra{h^{(i)}_{\rm min}} + \ket{h^{(i)}_{\rm min}}\bra{h^{(i)}_{\rm max}} \Big)_i, \nonumber
\eeq
and $\hat{\Pi}_1 = \Eins - \hat{\Pi}_0$, given by separate measurements on each subsystem.
We find $P(\pi_0\vert \theta) = \bra{\psi} e^{i \hat H \theta} \hat{\Pi}_0 e^{-i \hat H \theta} \ket{\psi} = \cos^2[N \theta (h_{\rm max}-h_{\rm min})/2]$
and $P(\pi_1\vert \theta) = \bra{\psi} e^{i \hat H \theta} \hat{\Pi}_1 e^{-i \hat H \theta} \ket{\psi} = \sin^2[N \theta (h_{\rm max}-h_{\rm min})/2]$, 
and FI 
\beq
F(\theta) = \frac{1}{P(\pi_0\vert \theta)} \Big( \frac{\ud P(\pi_0\vert \theta)}{\ud \theta}\Big)^2 + 
\frac{1}{P(\pi_1\vert \theta)} \Big( \frac{\ud P(\pi_1\vert \theta)}{\ud \theta}\Big)^2 = N^2(h_{\rm max} - h_{\rm min})^2. \nonumber
\eeq
} giving the right hand side of Eq.~(\ref{Fishentent}).
The quantum enhancement of phase sensitivity offered by the state (\ref{MaxEnt}), has been experimentally
verified with trapped ions \cite{LeibfriedSCIENCE2004, LeibfriedNATURE2005, MonzPRL2011} and 
photons \cite{WaltherNATURE2004, MitchellNATURE2004}, see also \cite{FonsecaPRL1999}
and implications in quantum imaging \cite{LugiatoJOB2002,TrepsPRL2002,BarnettEPJD2003,ShihIEEE2007} 
and quantum lithography \cite{BotoPRL2000, DAngeloPRL2001, EdamatsuPRL2002}.


\subsubsection{$k$-particle entangled probe}
\label{Sec:kentangled}

-- In the previous sections we have considered the two limit cases:
the fully separable state and the maximally entangled one, which saturate the shot noise and the Heisenberg limit, respectively.
For many-particle systems, it is interesting to consider the intermediate cases where 
only a fraction of the $N$ subsystems are in an entangled state.
Let us start from a definition.
A pure state of $N \geq 2$ particles is {\it $k$-producible} \cite{SeevinckPRA01, GuehneNJP2005, GuehnePRA2009, ChenPRA05, SorensenPRL2001}
if it can be written as a tensor product of the form 
\be \label{k-separable}
\ket{\psi_{k-{\rm prod}}} = 
\ket{\psi_1} \otimes \ket{\psi_2} \otimes ... \otimes \ket{\psi_M}, 
\ee
where $\ket{\psi_l}$ is a state of $N_l \leq k$
particles, with $\sum_{l=1}^M N_l=N$. 
This definition is straightforwardly extended to the case of mixed states:
a mixed state is $k$-producible if it can be
written as a mixture of $k_l$-producible pure states,
\be \label{k-mixedseparable}
\rho_{k-{\rm prod}}=\sum_l p_l \proj{\psi_{k_l-{\rm prod}}}, \quad \quad {\rm with}~k_l\le k,
\ee
where $p_l> 0$ and $\sum_l p_l=1$.
A state (pure or mixed) is {\it $k$-particle entangled} if it is 
$k$-producible but not $(k-1)$-producible.
In other words, a pure $k$-particle entangled state 
can be written as $\vert \psi_{k-{\rm ent}} \rangle=\otimes_{l=1}^M \ket{\psi_l}$, 
where the product contains at least one state $\ket{\psi_l}$ of $N_l=k$ particles which does not factorize~\footnote{
Let us illustrate the classification by considering states of $N=3$ particles. 
A state $\ket{\psi_{1-{\rm prod}}}=\ket{\phi_1}\otimes\ket{\varphi_2}\otimes\ket{\chi_2}$
is fully separable. 
A state $\ket{\psi_{2-{\rm ent}}}=\ket{\phi}_{12}\otimes\ket{\chi}_3$
which cannot be written as $\ket{\psi_{1-{\rm prod}}}$ 
(i.e. $\ket{\phi_{1,2}}$ do not factorize, $\ket{\phi_{1,2}} \neq \ket{\phi_1}\otimes\ket{\varphi_2}$) 
is $2$-particle entangled.
A state $\ket{\psi_{3-{\rm ent}}}$ which does not factorize at all is $3$-particle entangled. 
}.
Using another terminology, we can say that a k-particle entangled state has an
entangled depth \cite{SorensenPRL2001} larger than $(k-1)$.

Here we find the criteria for useful multiparticle entanglement \cite{HyllusPRA2012, TothPRA2012}. 
Starting from the definition~(\ref{k-mixedseparable}) and using the convexity of the QFI, we have 
\be
F_Q\big[\rho_{k-{\rm prod}} \big] 
\leq \sum_l p_l F_Q\big[ \ket{\psi_{k_l-{\rm prod}}} \big]
\leq \sum_l p_l 4 \big( \Delta \hat{H} \big)^2_{ \ket{\psi_{k_l-{\rm prod}}} }.
\ee
Since $\hat{H}$ is linear and $\ket{\psi_{k_l-{\rm prod}}}$ is the product (\ref{k-separable}), 
we have~\footnote{
It is easy to see that for a product state $\ket{\phi_A}\otimes\ket{\chi}_B$ and a linear Hamiltonian $\hat H_{AB} = \hat{H}_{A}\otimes \hat H_{B}$, 
we have $(\Delta\hat H_{AB})^2_{\ket{\phi}_A\otimes\ket{\chi}_B}
=(\Delta\hat H_{A})^2_{\ket{\phi}_A}+(\Delta\hat H_{B})^2_{\ket{\chi}_B}$.
Here $\hat H_{AB}$ acts on all the particles while $\hat H_{A}$ acts
on the particles of $\ket{\psi}_A$ only and in analogy for $\hat H_{B}$.}
\be
4\big( \Delta \hat{H} \big)^2_{ \ket{\psi_{k_l-{\rm prod}}} } 
= \sum_{l=1}^M 4 \big( \Delta \hat{H} \big)^2_{ \ket{\psi_{l}} }
\leq \sum_{l=1}^M \big( H_{\rm max}^{(l)} - H_{\rm min}^{(l)} \big)^2,
\ee
where $H_{\rm max}^{(l)}$ and $H_{\rm min}^{(l)}$ are the maximum and 
minimum eigenvalues of $\hat{H}^{(l)} = \otimes_{k=1}^{N_l} \hat{h}_i$, respectively,
and $\hat{h}^{(i)}$ is the single particle Hamiltonian.
In our case we have $H_{\rm max}^{(l)} = N_l h_{\rm max}$ and $H_{\rm min}^{(l)} = N_l h_{\rm min}$.
Putting all these results together
we find
\be
\max_{\rho_{k-{\rm prod}}} F\big[\rho_{k-{\rm prod}} \big] \leq (h_{\rm max}-h_{\rm min})^2 
\max_{\{N_l\}} \sum_{l=1}^M N_l^2,
\ee
where the maximum on the right hand side of this equation is calculated over all possible 
partitions $\{N_l\}$ of the system according to $\sum_{l=1}^{M} N_l = N$. 
Since $(N_1+1)^2+(N_2-1)^2\ge N_1^2 + N_2^2$ if $N_1\ge N_2$,
the right hand side of the equation above is increased by making the $N_l$ as
large as possible.
\begin{figure}[!t]
\begin{center}
\includegraphics[scale=0.5]{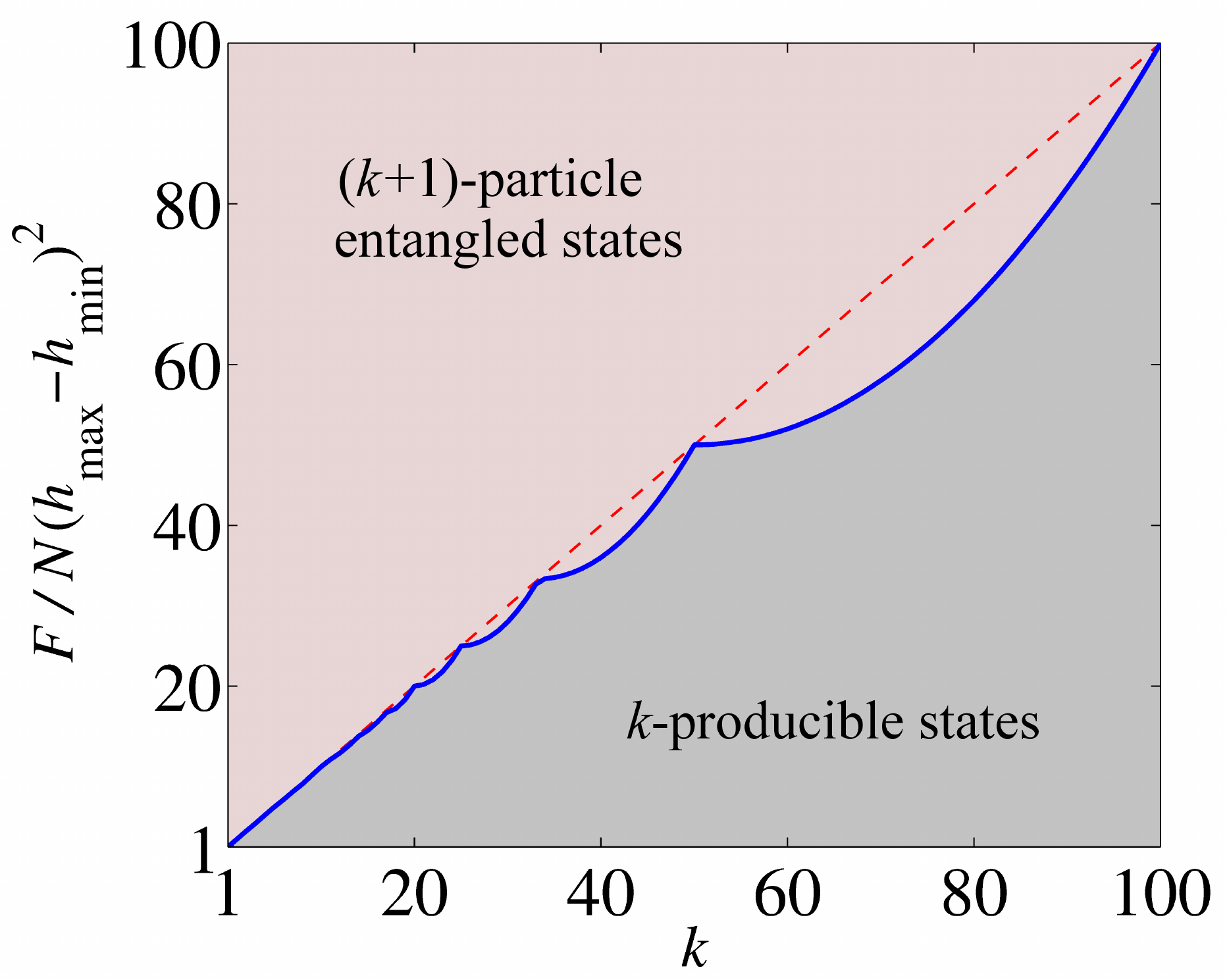}
\end{center}
\caption{\small{ {\bf Criterion for multiparticle entanglement from the Fisher information:} 
The solid line is the bound $F_Q/(h_{\rm max}-h_{\rm min})^2  = (s k^2 + r^2) $
which separates $k$-producible states (below the line)
from $(k+1)$-particle entangled states (above the line).
The linear behaviour $F_Q/(h_{\rm max}-h_{\rm min})^2=Nk$ is plotted for comparison (dashed line). 
Here $N=100$.}}
\label{Fig1} 
\end{figure}
For a $k$-producible state $N_l \leq k$ and therefore 
$\max_{\{N_l\}} \sum_{l=1}^M N_l^2 = s k^2 + r^2$, where 
$s=\lfloor \frac{N}{k}\rfloor$ is the largest integer smaller than
or equal to $\frac{N}{k}$ and $r=N-s k$.
The maximum FI is thus 
reached by the product of $s$ GHZ states of $k$ particles and a GHZ state with the remaining $r$ particles:
\beq
\ket{\psi} = 
\bigotimes_{i=1}^s \bigg( \frac{ \ket{h_{\rm max}}^{\otimes k} + \ket{h_{\rm min}}^{\otimes k} }{\sqrt{2}} \bigg)_i \otimes 
\bigg( \frac{ \ket{h_{\rm max}}^{\otimes r} + \ket{h_{\rm min}}^{\otimes r} }{\sqrt{2}} \bigg).
\eeq
Therefore, for $k$-producible states, we find the bound 
\be \label{FQ_class}
F\big[\rho_{k-{\rm prod}} \big] \le (h_{\rm max}-h_{\rm min})^2 \big( s k^2 + r^2 \big).
\ee
Given the linear operator $\hat H$ and the generic probe state $\hat \rho$, 
the criterion~(\ref{FQ_class}) has a clear operational meaning.
If the bound is surpassed, then the probe state contains useful $(k+1)$-particle entanglement:
when used as input state of the interferometer defined by the transformation 
$e^{-i\theta\hat H}$, $\hat \rho$ enables a phase sensitivity better than any $k$-producible state.
A plot of the bound Eq.~(\ref{FQ_class}) is presented in Fig.~\ref{Fig1} as a function of $k$. 
Since the bound increases monotonically with $k$, 
the maximum achievable phase sensitivity increases with
the number of entangled particles.
For $k=1$ we recover the bound~(\ref{Fishsepsep})
valid for separable states. For $k=N-1$, the bound is
$F[\rho_{(N-1)-{\rm prod}}]/(h_{\rm max}-h_{\rm min})^2 \le (N-1)^2+1$
and a QFI larger than this value
signals that the state is fully $N$-particle entangled.
The maximum value of the bound is obtained for $k=N$
(thus $s=1$ and $r=0$) giving $F_Q[\rho_{N-{\rm ent}}]/(h_{\rm max}-h_{\rm min})^2 = N^2$ and
thus recovering Eq.~(\ref{Fishentent}).

\section{SU(2) interferometry}
\label{SU2}

In this section we study phase estimation with a collection of $N$ qubits,
e.g. $N$ particles in two modes. 
For the single qubit, we consider the rotation  $e^{-i \theta \hat \sigma_{\vect{n}}}$,
where $\hat \sigma_{\vect{n}}$ the Pauli matrix, $\vect{n}$ is an arbitrary direction  in the Bloch sphere
and $\theta$ is the rotation angle.
The collective rotation of $N$ qubits (of the same angle $\theta$ and around the same axis $\vect{n}$) 
is given by the unitary operator
\be \label{UnitQ}
\hat{U}(\theta) = e^{-i \theta \hat J_{\vect{n}}},
\ee
where $\hat J_{\vect{n}} \equiv \sum_{i=1}^N \hat \sigma^{(i)}_{\vect{n}}$ and  
$\hat \sigma^{(i)}_{\vect{n}}$ is the Pauli matrix for the $i$th qubit.
This transformation rotates the pseudo-spin operator 
$\hat{\vect{J}} \equiv(\hat J_x, \hat J_y, \hat J_z)$ around the $\vect{n}$ axis in the generalised Bloch sphere.
The rotation angle $\theta$ is the parameter we want to estimate.
How to create in practice this model ? We will see that Eq.~(\ref{UnitQ})
can be implemented by linear two-mode 
atomic (e.g. Ramsey) and optical (e.g. Michelson and Mach-Zehnder) 
interferometers. Applications range from the detection of gravitational waves with laser interferometers
to the measurement of time, forces, gradient and accelerations with atoms.


\subsection{Collective two-mode transformations and Schwinger Formalism}

The general linear transformation of a two-mode system is 
\begin{equation} \label{BS}
{\hat{a}_{\rm out} \choose \hat{b}_{\rm out}}= \left( \begin{array}{cc}
m_{11} & m_{12} \\
m_{21} & m_{22} 
\end{array} \right)
{\hat{a}_{\rm in} \choose \hat{b}_{\rm in}},
\end{equation}
where $m_{ij}=|m_{ij}|e^{i \phi_{ij}}$ are complex numbers defining the scattering matrix  $\mathbf{M} =\{m_{ij} \}$
and $\hat{a}_{\rm in}$,$\hat{b}_{\rm in}$ and $\hat{a}_{\rm out}$, $\hat{b}_{\rm out}$ are 
the annihilator (creation) operators for the input and output modes, respectively.
Imposing the conservation of the total number of particles, 
$\hat{a}^{\dag}_{\rm in}\hat{a}_{\rm in} + \hat{b}^{\dag}_{\rm in}\hat{b}_{\rm in} = 
\hat{a}^{\dag}_{\rm out}\hat{a}_{\rm out} + \hat{b}^{\dag}_{\rm out}\hat{b}_{\rm out}$, we obtain three conditions:
\begin{eqnletter}
&& \vert m_{11} \vert^2 + \vert m_{21} \vert^2 =1, \\
&& \vert m_{22} \vert^2 + \vert m_{12} \vert^2 =1, \\
&& m_{11} m_{12}^* + m_{21} m_{22}^*=0.
\end{eqnletter}
The general scattering matrix that fulfils these requirements is 
\begin{equation} \label{BS1}
\mathbf{M} = e^{-i\phi_0} \, \left( \begin{array}{cc}
t & -r \\
r^\ast & t^\ast 
\end{array} \right),
\end{equation}
where $r$ and $t$ satisfy $|r|^2 + |t|^2 = 1$ and can 
be physically interpreted as Fresnel reflection and transmission coefficients, respectively.
Equation~(\ref{BS1}) is a unitary matrix with unit determinant, $\det \mathbf{M} = e^{-2 i \phi_0}$:
it is therefore the most general transformation of the U(2) group, where 
the unitarity stems from the conservation of the probability/total number of particles between the input and output ports.
It preserves Fermi and Bose commutation relations.
If we choose $\phi_0=0$, we restrict the transformation to the unimodular 
($\det \mathbf{M} = 1$) subgroup SU(2).
This restriction holds in the experimentally relevant situation when the input state and/or output POVM 
do not contain coherences between different number of particles \cite{HyllusPRL2010, JarzynaPRA2012}. 
By writing
\be \label{rt}
t = e^{-i \phi_{t}} \cos \frac{\vartheta}{2}, \qquad 
r = e^{-i \phi_{r}} \sin \frac{\vartheta}{2}, 
\ee
with $0 \le \vartheta \le \pi$ and $0 \le \phi_t, \phi_r \le 2\pi$,
the scattering matrix becomes
\begin{equation} \label{euler}
\mathbf{M}_{\rm SU(2)} = \left( \begin{array}{cc}
e^{-i \phi_t} \, \cos\frac{\vartheta}{2} & - e^{-i \phi_r} \, \sin\frac{\vartheta}{2} \\
e^{i \phi_r} \, \sin\frac{\vartheta}{2} & e^{i \phi_t} \, \cos \frac{\vartheta}{2} 
\end{array} \right).
\end{equation}
The elegant formalism developed by Schwinger in the 50's \cite{Schwinger, Biederharn} shows that 
the SU(2) group is equivalent to the SO(3) rotation group in three dimensions.
Therefore, we will identify the transformation (\ref{euler}) as a rotation of the vector 
\begin{equation}
\hat{\vect{J}} \equiv 
\left( \begin{array}{c}
\hat{J}_x \\
\hat{J}_y \\
\hat{J}_z \\
\end{array} \right)
=\frac{1}{2}
\left( \begin{array}{c}
\hat{a}^{\dag}\hat{b} + \hat{b}^{\dag}\hat{a} \\
 -i(\hat{a}^{\dag}\hat{b}-\hat{b}^{\dag}\hat{a})\\
\hat{a}^{\dag}\hat{a}-\hat{b}^{\dag}\hat{b}\\
\end{array} \right),
\end{equation}
mathematically analogous to the angular momentum,
in an abstract three-dimensional space. 
The connection between the SU(2) group and linear lossless quantum interferometry has been first recognized 
by Yurke \cite{Yurke_1986_b, Yurke_1986} and more recently reviewed by other authors
\cite{Luis_2000, Campos_1989, Kim_1997}.
The operator $\hat J_z$ is the relative number of particles operator among the two modes.
It sets a quantisation axis.
The three operators $\hat{J}_x$, $\hat{J}_y$ and $\hat{J}_z$ satisfy the commutation relations
\be
\big[\hat{J}_x, \hat{J}_y\big] = i \hat{J}_z, \qquad \big[\hat{J}_x, \hat{J}_z\big] = - i \hat{J}_y, \qquad \big[\hat{J}_y,\hat{J}_z\big]=i\hat{J}_z.
\ee
The Casimir invariant
\begin{equation}
\hat{\vect{J}}^2 = \hat{J}_{x}^2+\hat{J}_{y}^2+\hat{J}_{z}^2 =
\frac{\hat{N}}{2}\Big(\frac{\hat{N}}{2}+1\Big),
\end{equation}
depends on $\hat{N}=\hat{a}^{\dag}\hat{a}+\hat{b}^{\dag}\hat{b}$
and, since it commutes with $\hat{J}_i$, $[\hat{J}_i, \hat{J}^2]=0$, $i=x,y,z$, 
is an integral of motion.
The most general SU(2) transformation (\ref{euler}) corresponds to a spin rotation 
$e^{-i \theta \hat{J}_{\vect{n}}}$, where $\theta$ and the direction $\vect{n}$ depend on $\vartheta$, $\phi_t$ and $\phi_r$.
The edge of the angular momentum $\hat{\vect{J}}$ remains on a sphere (generally indicated as the generalised Bloch sphere), 
as a consequence of the lossless property of the unitary transformation.
This can be demonstrated by using the operator identity:
\begin{equation} \label{BCH}
e^{\delta \hat{A}} \hat{B} e^{-\delta \hat{A}} = 
\hat{B}+
\delta [\hat{A}, \hat{B}] +
\frac{\delta^2}{2!} [\hat{A}, [\hat{A},\hat{B}]]+ 
\frac{\delta^3}{3!} [\hat{A}, [\hat{A}, [\hat{A},\hat{B}]]] + ... .
\end{equation}
In the Heisenberg picture, the angular momentum components transform as
\be
\hat{\vect{J}}_{\rm out} = e^{+i \theta \hat{J}_n} \hat{\vect{J}}_{\rm in} e^{-i \theta \hat{J}_n}, 
\ee
while the equivalent evolution in Schr\"odinger representation of a two mode state is
\begin{equation}
|\psi_{\rm out}\rangle=e^{-i\theta\hat{J}_n}|\psi_{\rm in}\rangle.
\end{equation}
In the following we consider three examples:


\subsubsection{The phase shifter}
-- The phase shifter only translates the phase of each output mode with respect to the input one.
With reference to the general two-mode transformation Eq.~(\ref{euler}), this operation corresponds to  
$t = e^{-i \theta/2}$ and $r = 0$, giving  
\begin{equation} \label{PhaseShift}
\bf{PS} =  \left( \begin{array}{cc}
e^{-i\frac{\theta}{2}} & 0 \\
0 & e^{i\frac{\theta}{2}}
\end{array} \right).
\end{equation}
It can easily be seen that the phase shifter corresponds to a rotation of the pseudo-spin $\hat{\vect{J}}$ by
an angle $\theta$ around the $z$ axis:
\begin{equation}
\left( \begin{array}{c}
\hat{J}_x \\
\hat{J}_y \\
\hat{J}_z \\
\end{array} \right)_{\rm out}
=
\left( \begin{array}{ccc}
 \cos \theta & -\sin \theta & 0\\
\sin \theta & \cos \theta & 0\\
0 & 0 & 1 \\
\end{array} \right)
\left( \begin{array}{c}
\hat{J}_{x} \\
\hat{J}_{y} \\
\hat{J}_{z} \\
\end{array} \right)_{\rm in},
\end{equation}
and can be represented by the unitary operator 
$\hat{U}_{\rm PS}(\theta)=e^{-i \theta \hat{J}_z}$.


\subsubsection{The symmetric beam splitter}
-- A symmetric beam splitter has the same effect on a beam incident 
in port $a$ as on a beam incident in port $b$. 
In Eq.~(\ref{euler}) this symmetry implies 
$t = t^\ast$ and $r = -r^\ast$.
We find the conditions 
 $e^{i \phi_t} = e^{-i \phi_t}$, which leads to $\phi_{t} = 0, \pi$, and
$e^{i \phi_r} = - e^{-i \phi_r}$, which leads to $\phi_r = \pm \frac{\pi}{2}$.
Finally, the beam splitter matrix can be written as (taking, without loss of generality, $\phi_r=0$, $\phi_t=-\pi/2$):
\begin{equation} \label{bsm}
\bf{BS} =  \left( \begin{array}{cc}
\cos \frac{\theta}{2} & - i \sin \frac{\theta}{2}  \\
- i \sin \frac{\theta}{2} & \cos \frac{\theta}{2}   
\end{array} \right).
\end{equation}
The 50-50 (balanced) symmetric beam splitter is obtained by imposing the further condition:
$\vert r \vert = \vert t \vert$, i.e. $\theta=\pm \pi/2$,
giving
\begin{equation} \label{bsm1}
{\bf BS}_{50-50} =  \frac{1}{\sqrt{2}} \left( \begin{array}{cc}
1 & \pm i \\
\pm i & 1  
\end{array} \right).
\end{equation}
Using the transformation described by Eq.~(\ref{bsm}), we obtain, after some 
algebra,
\begin{equation} \label{BS_matrix}
\left( \begin{array}{c}
\hat{J}_x \\
\hat{J}_y \\
\hat{J}_z \\
\end{array} \right)_{\rm out}
=
\left( \begin{array}{ccc}
1 & 0 & 0 \\
0 &  \cos \theta & -\sin \theta \\
0 & \sin \theta & \cos \theta
\end{array} \right)
\left( \begin{array}{c}
\hat{J}_{x} \\
\hat{J}_{y} \\
\hat{J}_{z} \\
\end{array} \right)_{\rm in}.
\end{equation}
The beam splitter can be viewed geometrically as a
rotation of the vector $\hat{\vect{J}}$, around the $x$ axis of an angle $\theta$ and can be 
represented by the unitary operator $\hat{U}_{\rm BS}(\theta)=e^{-i \theta \hat{J}_x}$. 


\begin{figure}[t!]
\begin{center}
\includegraphics[clip,scale=0.28]{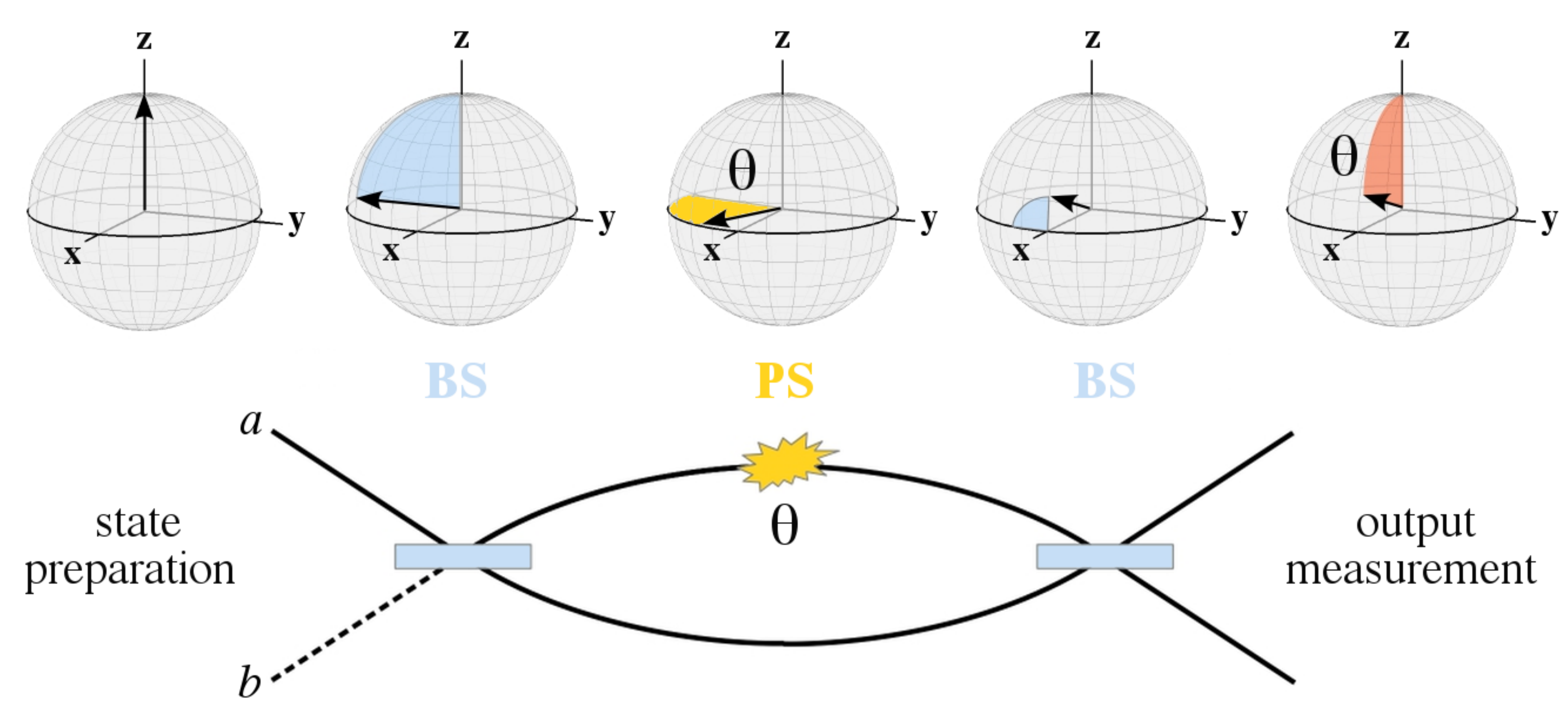}
\end{center}
\caption{
Schematic diagrams of the Ramsey (above) and Mach-Zehnder (below) interferometers 
with input state given by $N$ particles in mode $a$ and zero particles in mode $b$, 
such that $\langle \hat{\vect{J}} \rangle = (0,0,N)$.
For the Ramsey interferometer, the black arrows point toward the
mean spin $\langle \hat{\vect{J}} \rangle$. 
For the Mach-Zehnder, the black lines represent the spatial path travelled by the light
(the dashed line in the input mode $b$ signifies that no photon enters that port).
In the Ramsey interferometer, the spin is first rotated by a $\pi/2$ angle around the $x$ axis ($\pi/2$ pulse) spanning the blue area,
it then precesses of an angle $\theta$ along the equator (yellow area) and it is finally rotated by a second $\pi/2$ pulse. 
The whole process is equivalent to a spin rotation of an angle $\theta$ around the $y$ axis (red area). 
In the Mach-Zehnder interferometer, the two beams mix at a 50-50 beam splitter (blue square), 
acquire a phase shift $\theta$, and mix again at a second 50-50 beam splitter. 
Typically, in both configurations, the phase shift $\theta$ is estimated from the field intensity at the two output modes 
(which can be visualised as a projection over the $z$ axis in the generalised Bloch sphere).  
} \label{Fig:RMZ}
\end{figure}

\subsubsection{Mach-Zehnder and Ramsey rotation}
-- The lossless balanced Mach-Zehnder interferometer is given by the sequence of a first 50-50 beam splitter, a  
phase shifter and a second 50-50 beam splitter (see Fig.~\ref{Fig:RMZ}): 
\begin{eqnarray} \label{MZ_matrix}
\frac{1}{\sqrt{2}}
\left( \begin{array}{cc}
1 & +i  \\
+i &  1 
\end{array} \right) \times
\left( \begin{array}{cc}
e^{-i\frac{\theta}{2}} & 0 \\
0 & e^{i\frac{\theta}{2}}
\end{array} \right) \times 
\frac{1}{\sqrt{2}} 
\left( \begin{array}{cc}
1 & -i  \\
-i &  1 
\end{array} \right)
& = &
\left( \begin{array}{cc}
\cos\frac{\phi}{2} & -\sin\frac{\phi}{2}  \\
\sin\frac{\phi}{2} &  \cos\frac{\phi}{2} 
\end{array} \right). \nonumber \\
\end{eqnarray}
This SU(2) transformation can be represented as
\begin{equation}
\left( \begin{array}{c}
\hat{J}_{x} \\
\hat{J}_{y} \\
\hat{J}_{z} \\
\end{array} \right)_{\rm out}
=
\left( \begin{array}{ccc}
\cos \theta & 0 & \sin \theta \\
0 & 1 & 0 \\
-\sin \theta & 0 & \cos \theta \\
\end{array} \right)
\left( \begin{array}{c}
\hat{J}_{x} \\
\hat{J}_{y} \\
\hat{J}_{z} \\
\end{array} \right)_{\rm in},
\end{equation}
which is a rotation of $\hat{\vect{J}}$ around the $y$ axis of an angle $\theta$. 
We can immediately recognize this by rewriting Eq.~(\ref{MZ_matrix})
in terms of rotation matrices and using $e^{i \frac{\pi}{2} \hat{J}_x} \hat{J}_z e^{-i \frac{\pi}{2} \hat{J}_x}=\hat{J}_y$:
\begin{equation} \label{MZ_trans}
\hat{U}_{\rm MZ}(\theta) = e^{i \frac{\pi}{2} \hat{J}_x} e^{-i \theta \hat{J}_z} e^{-i \frac{\pi}{2} \hat{J}_x} = e^{-i \theta \hat{J}_y}.
\end{equation}
In Eqs.~(\ref{MZ_matrix})-(\ref{MZ_trans}), we made the arbitrary assumption that
the two beam splitters rotate of opposite angles. 
If the beam splitters rotate of the same angle, we obtain $e^{i \frac{\pi}{2} \hat{J}_x} e^{-i \theta \hat{J}_z} e^{i \frac{\pi}{2} \hat{J}_x}=
e^{-i \theta \hat{J}_y} e^{i \pi \hat{J}_x}$ which is equivalent to Eq.~(\ref{MZ_trans}) modulo a rotation of the probe state.
We finally point out that Ramsey spectroscopy and Mach-Zehnder interferometery are formally 
equivalent and described by the above equations \cite{WinelandPRA1994, LeeJMO2002}. 
In this analogy (see Fig.~\ref{Fig:RMZ}), 50-50 beam splitters are equivalent to $\pi/2$ pulses
and the phase shift accumulated during the spin precession between $\pi/2$ pulses
corresponds to the relative phase shift between the arms of a Mach-Zehnder interferometer.
In Ramsey spectroscopy the two modes supporting the dynamics are 
the two internal levels of an atom \cite{RamseyBOOK} and can eventually be coupled 
to external motional degrees of freedom \cite{BorderJPA1989}.
In Mach-Zehnder interferometry the two modes are the 
two spatially-separated arms.    


\subsection{Symmetric subspace and rotation matrix elements}

Collective qubit transformations as in Eq.~(\ref{UnitQ}) allow for a crucial simplification:  
while the Hilbert space of $N$ qubits has a dimension $2^N$, the operators $\hat{J}_{\vect{n}}$
can be fully diagonalised in the permutationally symmetric $(N+1)$-dimensional subspace. 
Without any loss of generality, this is most conveniently seen along the $z$ axis.
We have 
\begin{equation} \label{diagJz}
\hat{J}_z \, |j, \mu \rangle_z = \mu |j, \mu \rangle_z,
\end{equation}  
where $j=N/2$ and $\mu=-j, -j+1,\ldots,j$ are $2j+1$ eigenvalues \footnote{
The linear spectrum is common to all operators  $\hat{J}_{\vect{n}}$. To find 
the corresponding eigenvectors one has to apply a proper rotation to $\vert j, \mu \rangle_z$, around an axis  
perpendicular to $\vect{n}$ and $z$, of an angle $\theta=\arccos(\vect{n} \cdot z)$.
For instance, $\vert j, \mu \rangle_x = e^{i \hat{J_y} \pi/2} \vert j, \mu \rangle_z$ and 
$\vert j, \mu \rangle_y = e^{-i \hat{J_x} \pi/2} \vert j, \mu \rangle_z$.
Indeed, using Eq.~(\ref{BS_matrix}), 
we have $\hat{J}_z = e^{i \hat{J_x} \pi/2} \hat J_{y} e^{-i \hat{J_x} \pi/2}$ and
Eq.~(\ref{diagJz}) becomes $\hat J_{y} e^{-i \hat{J_x} \pi/2} \, |j, \mu \rangle_z = \mu \, e^{-i \hat{J_x} \pi/2} |j, \mu \rangle_z$, 
showing  that $e^{-i \hat{J_x} \pi/2} \, |j, \mu \rangle_z$ is the eigenstate of $\hat J_{y}$ with eigenvalue $\mu$.
} with corresponding eigenvectors 
\begin{eqnletter}
|j, \mu \rangle_z &\equiv& |j+\mu\rangle_a |j-\mu\rangle_b, \\
&=& \frac{1}{\sqrt{\binom{2j}{j+\mu}}} \,
 \mathcal{S} \Big[ \vert \uparrow \rangle_1 \ldots \vert \uparrow \rangle_{j+\mu} \vert \downarrow \rangle_{j+\mu+1} \ldots \vert \downarrow \rangle_{2j} \Big], \label{permeq}
\end{eqnletter}
given by a state 
with $j+\mu$ particles in mode $a$ and $j-\mu$ particles in mode $b$. 
According to Eq.~(\ref{permeq}), this state is obtained by symmetrizing the single-particles states (labelled by $1,2,\ldots,2j$), 
where $\mathcal{S}$ represents the sum 
of all permutations of $2j$ particles among which $j+\mu$ have pseudo spin up ($\vert \uparrow \rangle$ or, equivalently, in mode $a$, with $\hat{\sigma}_z \vert \uparrow \rangle = \frac{1}{2}  \vert \uparrow \rangle$) and the remaining have pseudo spin down ($\vert \downarrow \rangle$ or, equivalently, in mode $b$, with $\hat{\sigma}_z \vert \downarrow \rangle = -\frac{1}{2}  \vert \downarrow \rangle$).
The states $|j, \mu \rangle_z$ are generally indicated as two-mode Fock states.
Physically, the restriction to the relevant particle-symmetrized Hilbert subspace is a consequence of the collective
nature of the qubit rotations. 

The operators $\hat{J}_x$ and $\hat{J}_y$ can be rewritten in terms of raising and lowering angular momentum 
operators, $\hat{J}_{\pm}=\hat{J}_x \pm i \hat{J}_y$ 
($\hat{J}_+=\hat{a}^{\dag}\hat{b}$ and $\hat{J}_-=\hat{b}^{\dag}\hat{a}$), acting
on the vector $|j, \mu \rangle_z$ as 
\be
\hat{J}_{\pm}|j, \mu \rangle_z = \sqrt{j(j+1)-\mu(\mu\pm1)} \, \vert j, \mu \pm 1 \rangle_z.
\ee
The elements of the rotation matrix $e^{-i\theta\hat{J}_{y}}$, 
$d^{j}_{\mu, \nu}(\theta) \equiv {_z}\langle j, \mu | e^{-i\theta\hat{J}_{y}} | j, \nu \rangle_z$,
are very useful in many calculations. An explicit expression of $d^{j}_{\mu, \nu}(\theta)$ is \cite{SandersPRL1995, Biederharn} 
\begin{eqnarray} \label{d.1}
d^{j}_{\mu, \nu}(\theta) =   \sqrt{\frac{(j-\nu)!(j+\nu)!}{(j-\mu)!(j+\mu)!}}
\Big( \sin\frac{\theta}{2} \Big)^{\nu-\mu}
\Big( \cos\frac{\theta}{2} \Big)^{\nu+\mu}
P_{j-\nu}^{\nu-\mu, \nu+\mu}(\cos \theta), 
\end{eqnarray}
being $P_{n}^{\alpha, \beta}(x)$ the Jacobi Polynomials \cite{AbramowitzBOOK}.
The rotation matrix elements~(\ref{d.1}) are real and satisfy the useful relations
\begin{eqnletter}
& & d^{j}_{\mu, \nu}(\theta)  = d^{j}_{\nu, \mu}(-\theta), \\
& & d^{j}_{\mu, \nu}(-\theta) = (-1)^{\mu-\nu} \, d^{j}_{\mu, \nu}(\theta), \\
& & d^{j}_{\mu, \nu}(\theta)  = (-1)^{\mu-\nu} \, d^{j}_{-\mu, -\nu}(\theta).
\end{eqnletter}
We also notice that 
\begin{eqnarray} \label{d.2}
{_z}\langle j, \mu | e^{-i \theta \hat{J}_{x}} | j, \nu \rangle_z =
{_z}\langle j, \mu | e^{-i \frac{\pi}{2} \hat{J}_{z}} e^{-i \theta\hat{J}_{y}} e^{i\frac{\pi}{2}\hat{J}_{z}}| j, \nu \rangle_z= e^{-i\frac{\pi}{2}(\mu-\nu)} d^{j}_{\mu, \nu}(\theta). 
\end{eqnarray}
More generally, using Eq.~(\ref{d.1}) it is possible to obtain an explicit expression for the rotation matrix 
around an arbitrary axis.


\subsection{Entanglement and phase sensitivity in SU(2) interferometry}

In this section we discuss the phase sensitivity of SU(2) interferometers
and its relation to particle (qubit) entanglement. 
We can immediately generalise the results of Sec.~\ref{QI&ENT} to the 
qubit case ($h_{\rm max} - h_{\rm min}=1$). 
We have two key bounds of phase sensitivity in linear SU(2) interferometers:
the shot noise limit, 
\be \label{SU2SN}
\Delta \theta_{\rm SN} = \frac{1}{\sqrt{m N}},
\ee
which is maximum sensitivity achievable with separable states, and 
the Heisenberg limit,
\be \label{SU2HL}
\Delta \theta_{\rm HL} = \frac{1}{N\sqrt{m}},
\ee
which is the maximum possible phase sensitivity allowed by quantum mechanics.
Let us discuss the saturation of these bounds with the measurement 
of the number of particles at the output ports of the interferometer. 
This measurement corresponds to projectors $\hat{\Pi}(\mu)\equiv \vert j, \mu \rangle_z \langle j, \mu \vert$
on eigenstates of $\hat{J}_z$.
The shot noise can be saturated, for instance, 
by the ``spin-polarized" state $\vert N\rangle_a \vert 0 \rangle_b$ rotated 
around the $y$ axis: $F\big[\vert N\rangle_a \vert 0 \rangle_b, \hat{J}_y, \{ \hat{\Pi}(\mu) \}\big] = N$ \cite{Yurke_1986}.
More generally, Eq.~(\ref{SU2SN}) is the relevant bound for the whole class of coherent spin states \cite{ArecchiPRA1972}.
The Heisenberg limit, Eq.~(\ref{SU2HL}), can be saturated (only) by the NOON state \cite{BollingerPRA1996}
\be
\vert {\rm NOON} \rangle = \frac{\vert N \rangle_a \vert 0 \rangle_b + \vert 0 \rangle_a \vert N \rangle_b}{\sqrt{2}}.
\ee
Besides the NOON state, there are other states that can provide a phase sensitivity scaling 
at the Heisenberg limit, $\Delta \theta \propto 1/N$, yet with a prefactor larger than 1.
A relevant example is the Twin Fock state \cite{HollandPRL1993}
\be \label{TwinF}
\vert {\rm TF} \rangle = \vert N/2 \rangle_a \vert N/2 \rangle_b,
\ee
which has $F\big[\vert N\rangle_a \vert 0 \rangle_b, \hat{J}_y, \{ \hat{\Pi}(\mu) \}\big] = N^2/2 +N$ for the usual output measurement of the particle number.

Finally, the condition of useful entanglement in SU(2) interferometry reads
\be \label{useful}
F\big[\hat \rho, \hat J_{\vect{n}}, \{ \POVM \} \big] >N.
\ee
To be more precise, Eq.~(\ref{useful}) means that, if we have an interferometer $e^{-i \theta \hat J_{\vect{n}}}$, an
input state $\hat \rho$ and a POVM $\{ \POVM \}$ -- such that the probability distribution of possible outcomes of the interferometer 
is $P(\varepsilon \vert \theta) = \tr[\POVM e^{-i \theta \hat J_{\vect{n}}} \hat \rho e^{+i \theta \hat J_{\vect{n}}}]$ --
an efficient estimation of $\theta$ has a sensitivity overcoming the shot noise Eq.~(\ref{SU2SN}). 
The condition~(\ref{useful}) cannot be fulfilled by any separable state.

\subsubsection{Quantum Fisher information}
-- Here we focus on the optimal phase sensitivity (optimised over all possible POVMs) that can be reached  with the generic two-mode probe state $\hat{\rho}$, 
rotated by unitary transformations $e^{-i \theta \hat J_{\vect{n}} }$. This is given by 
the quantum Cramer-Rao bound $\Delta \Theta_{\rm QCR} = 1/\sqrt{m F_Q[\hat \rho, \hat J_{\vect{n}}]}$, where the QFI is (see Sec.~\ref{unitaryQFI})
\begin{equation} \label{SU2QFIps}
F_Q\big[\vert \psi \rangle, \hat J_{\vect{n}}\big] = 4\big( \Delta \hat J_{\vect{n}} \big)^2,
\ee
for pure states, and 
\begin{equation} \label{SU2QFIrho}
F_Q\big[\hat \rho, \hat J_{\vect{n}}\big] = 2 \sum_{k,k'} \frac{ (p_{k} - p_{k'})^2}{ p_k + p_{k'} } \, \big\vert \bra{k} \hat J_{\vect{n}} \ket{k'} \big\vert^2.
\ee
for mixed states $\hat{\rho} = \sum_k p_k \ket{k}\bra{k}$ (where $\{ \vert k \rangle \}$ is an orthogonal basis set, $p_k\geq 0$, $\sum_k p_k=1$ and the 
sum in Eq.~(\ref{SU2QFIrho}) extends over $p_k + p_{k'} \neq 0$).
We recall that $F_Q\big[\hat \rho, \hat J_{\vect{n}} \big] \leq 4\big( \Delta \hat J_{\vect{n}} \big)^2$, where the inequality is not tight, in general.
It should be noticed that Eq.~(\ref{useful}) depends on the specific POVM considered: 
a state that is useful with respect to a certain POVM may not be useful if a different POVM is chosen.  
For this reason we can thus give a further condition of useful entanglement based on the QFI:
\be \label{useful2}
F_Q\big[\hat \rho, \hat J_{\vect{n}} \big] >N.
\ee
If Eq.~(\ref{useful2}) is fulfilled, there is at least one optimal POVM (e.g. the POVM for which $F=F_Q$) such that 
$F[\hat \rho, \hat J_{\vect{n}}, \{ \POVM \} ] >N$. 
On the other hand, Eq.~(\ref{useful}) implies Eq.(\ref{useful2}).

An interesting situation is to find the spin direction, in the Bloch sphere,
for which the FI reaches its maximum value, given a pure or mixed state \cite{HyllusPRA2010}. 
For pure states, this problem can be solved by noticing that $\hat{J}_{\vect{n}} = \vect{n} \cdot \hat{\vect{J}}$ and thus
the variance $4(\Delta \hat J_{\vect{n}})^2 = 4 \vect{n}^T \langle ( \vect{\hat{J}} - \langle \vect{\hat{J}} \rangle) ( \vect{\hat{J}} - \langle \vect{\hat{J}} \rangle) \rangle \vect{n}$
can be written in terms of the $3\times 3$ covariance matrix.
Since only the real part is relevant, we have 
$F_Q \big[\ket{\psi}, \hat J_{\vect{n}} \big]= 4 \vect{n}^T \,  \vect{\gamma}_C \, \vect{n}$, where $\vect{\gamma}_C$ is a real matrix of entries
\begin{equation}
[\vect{\gamma}_C]_{i,j} =  \frac{\langle \hat J_i \hat J_j \rangle + \langle \hat J_j \hat J_i \rangle}{2}  - \langle \hat J_i \rangle \langle \hat J_j \rangle.
\ee
It is known from linear algebra that this expression is maximized by choosing 
$\vect{n} = \vect{n}_{\rm max}$ as the eigenvector corresponding to the maximum eigenvalue
$\lambda_{\rm max}$.
The QFI maximized over all possible directions $\vect{n}$ is thus
given by $4\lambda_{\rm max}$, and the optimal direction given by $\vect{n}_{\rm max}$.
For a mixed state $\hat \rho$, the analogous maximisation can be obtained by noticing that $\vert \bra{k} \hat J_{\vect{n}} \ket{k'} \vert^2 = 
(\vect{n} \cdot \bra{k'} \hat{\vect{J}} \ket{k}) (\vect{n} \cdot \bra{k} \hat{\vect{J}} \ket{k'}) = 
\vect{n}^T \bra{k'} \hat{\vect{J}} \ket{k}  \bra{k} \hat{\vect{J}} \ket{k'} \, \vect{n}$, so we obtain
that $F_Q\big[\hat J_{\vect{n}}, \hat \rho \big]= 4 \vect{n}^T \,  \vect{\Gamma}_C \, \vect{n}$, 
where $\vect{\Gamma}_C$ is a real matrix of entries
\begin{equation}
[\vect{\Gamma}_C]_{i,j} = \frac{1}{2} \sum_{k,k'} \frac{(p_{k} - p_{k'})^2}{ p_k + p_{k'} }
\bra{k'} \hat{J}_i \ket{k}  \bra{k} \hat{J}_j \ket{k'}.
\ee
The maximum QFI is then obtained as four times the maximum eigenvalue
of the matrix $\vect{\Gamma}_C$ and the optimal direction is given by the corresponding eigenvector.


\subsubsection{Spin squeezing}
-- The method of moments discussed in Sec.~\ref{MethodMom} can be applied for phase estimation in SU(2)
interferometry \cite{WinelandPRA1992, WinelandPRA1994}. 
The most straightforward choice of observable is the spin operator along a direction $\vect{n}_1$ orthogonal 
to the rotation direction $\vect{n}_2$ (we indicate with $\vect{n}_1$, $\vect{n}_2$ and $\vect{n}_3$ there orthogonal directions
in the Bloch sphere and $e^{-i \theta \hat{J}_{\vect{n}_2}}$ the interferometer transformation).
The method of moments predicts 
\be \label{SU2errprop}
(\Delta \theta)^2 = \frac{(\Delta \hat{J}_{\vect{n}_1})^2}{ m \vert \partial \mean{\hat{J}_{\vect{n}_1}}/\partial \theta \vert^2}, \quad \quad {\rm for}\,\, m\gg 1.
\ee
Using Eq.~(\ref{BCH}) and $[\hat{J}_{\vect{n}_1}, \hat{J}_{\vect{n}_2}] = i \hat{J}_{\vect{n}_3}$,
we have $\mean{\hat{J}_{\vect{n}_1}} = \mean{\hat{J}_{\vect{n}_1}} \cos \theta + \mean{\hat{J}_{\vect{n}_3}} \sin \theta$.
The phase sensitivity (\ref{SU2errprop}) calculated at $\theta=0$ is 
\be
(\Delta \theta)^2 = \frac{(\Delta \hat{J}_{\vect{n}_1})^2}{ m \mean{\hat{J}_{\vect{n}_3}}^2},  \quad \quad {\rm for}\,\, m\gg 1.
\ee
Using this equation, we can rewrite the condition for sub shot noise phase sensitivity:
\be \label{Eq:xiWineland}
\xi_R^2 \equiv \frac{(\Delta \theta)^2}{(\Delta \theta)^2_{\rm SN}} = \frac{ N (\Delta \hat{J}_{\vect{n}_1})^2 }{\mean{\hat{J}_{\vect{n}_3}}^2} < 1,
\ee
known as spin squeezing condition, first introduced in Ref.~\cite{WinelandPRA1992, WinelandPRA1994}.
If the inequality $\xi_R^2<1$ holds, the state is said to be spin squeezed along the direction $\vect{n}_1$ 
\cite{WinelandPRA1992, WinelandPRA1994,SorensenNATURE2001,MaPHYSREP2012}.
In the literature, different definitions of spin squeezing for pseudo angular momentum operators can be found
\cite{WinelandPRA1992, WinelandPRA1994, SorensenNATURE2001, WallsPRL1981, KitagawaPRA1993}
(for a review, see~\cite{MaPHYSREP2012}), nevertheless Eq.~(\ref{Eq:xiWineland}) is the one directly related 
to interferometric sensitivity.

We already know, from the results of Sec.~\ref{QI&ENT}, that 
sub shot noise cannot be surpassed by separable state. 
Equation~(\ref{Eq:xiWineland}) is thus a sufficient condition for entanglement. 
We can demonstrate this directly by using the properties of the spin operators.
More precisely, we demonstrate that, for separable states, the following inequality holds~\cite{SorensenNATURE2001}:
\be \label{Eq:xiSorensen}
\xi_{R'}^2 = \frac{N (\Delta \hat{J}_{\vect{n}_1})^2}{ \mean{\hat{J}_{\vect{n}_2}}^2+\mean{\hat{J}_{\vect{n}_3}}^2} \geq 1.
\ee
A violation of the inequality $\xi_{R'}^2 \geq 1$ implies that the state is entangled.
Note that the denominator in Eq.~(\ref{Eq:xiSorensen})
gives the spin length on the plane perpendicular to $\vect{n}_1$.
Therefore, the spin squeezing conditions $\xi_{R'}^2 < 1$ and $\xi_{R}^2 < 1$ are equivalent, modulo a rotation around the $\vect{n}_1$ axis
(which does not change the entanglement properties of the state). States violating Eq.~(\ref{Eq:xiSorensen}) are thus also useful for sub shot noise phase estimation.

The inequality (\ref{Eq:xiSorensen}) can be obtained by first noticing that the variance 
$(\Delta \hat{A})^2 = \mean{\hat{A}^2} - \mean{\hat{A}}^2$ of any operator $\hat{A}$ is concave in the state:
if $\hat{\rho} = p \hat{\rho}_1 + (1-p) \hat{\rho}_2$, we have 
$(\Delta \hat{A})_{\hat{\rho}}^2 \geq p (\Delta \hat{A})_{\hat{\rho}_1}^2 + (1-p) (\Delta \hat{A})_{\hat{\rho}_2}^2$, for any $0\leq p \leq 1$.
We can thus restrict the demonstration of Eq.~(\ref{Eq:xiSorensen}) to pure (separable) states.
Equation (\ref{Eq:xiSorensen}) follows from three simple facts:
{\it i}) the inequality
\be \label{Pauli}
\mean{ \hat{\sigma}_{\vect{n}_1} }^2 + 
\mean{ \hat{\sigma}_{\vect{n}_2} }^2 +
\mean{ \hat{\sigma}_{\vect{n}_3} }^2 \leq \frac{1}{4}, 
\ee 
valid for the single-qubit;
{\it ii}) the inequality
\be \label{CS}
\bigg( \sum_{i=1}^{N} s_i \bigg)^2  \leq N \, \sum_{i=1}^{N} s_i^2,
\ee
where $s_i$ are arbitrary real numbers;
{\it iii}) the basic property $\mean{ \hat{\sigma}_{\vect{n}}^{(i)} \hat{\sigma}_{\vect{n}}^{(j)} } = 
\mean{ \hat{\sigma}_{\vect{n}}^{(i)} } \mean{ \hat{\sigma}_{\vect{n}}^{(j)} }$, for $i \neq j$,
which holds for separable states and, in particular, implies 
(recalling that $\langle \sigma_{\vect{n}}^2 \rangle= 1/4$ and 
$\hat{J}_{\vect{n}} =  \sum_{i=1}^N \hat{\sigma}_{\vect{n}}^{(i)} $)
\be \label{Var}
(\Delta \hat{J}_{\vect{n}})^2 = \frac{N}{4} - \sum_{i=1}^N \mean{ \hat{\sigma}_{\vect{n}}^{(i)} }^2.
\ee
To demonstrate Eq.~(\ref{Eq:xiSorensen}) we first use
Eq.~(\ref{Var}) and then the inequality~(\ref{Pauli}) to obtain
\be
(\Delta \hat{J}_{\vect{n}_1})^2 \geq \sum_{i=1}^N \mean{ \hat{\sigma}_{\vect{n}_2}^{(i)} }^2 + 
\sum_{i=1}^N \mean{ \hat{\sigma}_{\vect{n}_3}^{(i)} }^2
\geq \frac{  \mean{\hat{J}_{\vect{n}_2} }^2 + \mean{\hat{J}_{\vect{n}_3} }^2 }{N},
\ee
where the last inequality follows from Eq.~(\ref{CS}).

Finally, we recall that it is possible to find a complete series of inequalities based on the first moments of 
the spin operator (one of those inequalities is $\xi_{R'}^2<1$) whose violation signals that the state is entangled, see Ref.~\cite{TothPRL2007}. 


\subsection{Spin squeezing and Fisher Information}

In section Sec.~\ref{bounds} we have provided a lower bound to the FI in terms of average moments of an arbitrary diagonal operator. 
We can easily adapt the bound (\ref{MomentIneq1}) to demonstrate that~\cite{PezzePRL2009}~\footnote{
We use Eq.~(\ref{MomentIneq1}) with $\hat{H}=\hat{J}_{\vect{n}}$ (i.e. we consider the phase shift operation $e^{-i \hat{J}_{\vect{n}}\theta}$), 
$\hat{M}=\hat{J}_{\vect{n}_1}$ and thus $\vert \langle [\hat M, \hat H ]\rangle \vert^2 = \langle \hat J_{\vect{n}_3} \rangle^2$, where
$\vect{n}_1$, $\vect{n}\equiv \vect{n}_2$ and $\vect{n}_3$ are three orthogonal directions. We thus find 
$F_Q [\hat \rho, \hat{J}_{\vect{n}}]\geq F \geq \langle \hat J_{\vect{n}_3} \rangle^2/(\Delta \hat{J}_{\vect{n}_1})^2=N/\xi_R^2$, 
with $\xi_R^2$ defined in Eq.~(\ref{Eq:xiWineland}), and recover Eq.~(\ref{FishVsSS}).
}
\be \label{FishVsSS}
\frac{N}{F_Q[\hat \rho, \hat{J}_{\vect{n}}]} \leq \xi_R^2. 
\ee
This inequality shows that if a state is spin squeezed, $\xi_R^2<1$ (along a direction orthogonal to the rotation direction $\vect{n}$), it also satisfies the condition of useful entanglement $F_Q[\hat \rho, \hat{J}_{\vect{n}}] > N$.
Since spin squeezing implies sub shot noise sensitivity, see Eq.~(\ref{Eq:xiWineland}), this results is certainly expected.
The contrary is not true: there are states which are not spin squeezed and yet usefully entangled \cite{StrobelSCIENCE2014}, 
the NOON state is an example.
 Note also that the results of section~\ref{bounds} are more general and apply, for instance, to 
the parity operator \cite{GerryCONTPHYS2010,KimPRA2013} and higher 
spin moments \cite{LueckeSCIENCE2011,KimPRA1998}. 

\section{Conclusions}

Quantum enhanced interferometry has been investigated in several proof-of-principle 
experiments in different optical and atomic systems. 
On the theory side, as reviewed in this paper, it leads us to a profound understanding about the role played by quantum 
correlations to overcome classical sensitivity limits.
Important recent developments, which for space reason have not been discussed in this Review,
include the robustness of quantum interferometers with respect to decoherence  \cite{HuelgaPRL1997,ShajiPRA2007,LiPRL2008,SinatraPRL2011,
FerriniPRA2011, EscherNATPHYS2011, BraunNATCOMM2011,DemkowiczNATCOMM2012,AcinPRL2012,DornerNJP2012,SzankowskiEPL2013,TothPRA2013}
the sensitivity when employing nonlinear phase-encoding transformations \cite{LuisPLA2004,BoixoPRL2007,RoyPRL2008,BoixoPRA2009,ZwierzPRL2010,NapolitanoNJP2010,NapolitanoNATURE2011}, 
the ultimate phase sensitivity bounds for states of a fluctuating number of particles
\cite{HyllusPRL2010, ZwierzPRL2010, ShapiroPRL1989,SummyOPTCOMM1990,ShapiroPRA1991,BraunsteinPRL1992,HallJMO1993,HradilPRA1995,OuPRL1996,HofmannPRA2009,
AnisimovPRL2010,HallPRA2012,HallNJP2012,RivasNJP2012,TsangPRL2012,GiovannettiPRL2012,GiovannettiPRL2012b,PezzePRA2013}
and multi-phase estimation \cite{DArianoPRA1997,GillPRA2000,MatsumotoJPA2002,MacchiavelloPRA2003,BallesterPRA2004,HumphreysPRL2013,
SpagnoloSCIREP2012,ZhangARXIV}.
We expect, in the near future, further crucial experimental and theoretical advancement on quantum interferometry aimed to transform the early results and prototypes 
in real-world technological applications. This would revolutionise the field of precision measurement. 

\acknowledgments
We thank all of our interferometric
colleagues for enlightening discussions over the recent years. 
Also, we thank Jan Chwede\'nczuk, Phillip Hyllus,
Francesco Piazza and Geza T\`oth for comments and discussions on this manuscript.
This work was supported by the European Commission small or medium-scale focused research project QIBEC 
(Quantum Interferometry with Bose-Einstein condensates, Contract Nr. 284584). 
L.P. acknowledges financial support by MIUR through FIRB Project No. RBFR08H058.

\end{document}